\documentclass[aps,prd,preprint,11pt, onecolumn, tightenlines, notitlepage, superscriptaddress, nofootinbib, preprintnumbers, floatfix]{revtex4}
\pdfoutput=1

\usepackage[x11names]{xcolor}
\usepackage[colorlinks]{hyperref}

\usepackage{amstext}
\usepackage{amssymb}
\usepackage{amsmath}
\usepackage{graphicx}
\graphicspath{{plots/}}
\usepackage{hyperref}
\usepackage{url}
\usepackage{color}
\usepackage[normalem]{ulem}
\usepackage[utf8]{inputenc}
\usepackage{textcomp}
\usepackage{todonotes}

\def\21{$\mathrm{SU(2)_L \otimes U(1)_Y}$}

\newcommand {\ignore}[1]{}
\usepackage{epsfig,amsfonts,mathrsfs,slashed,multirow}
\usepackage{latexsym,enumerate,cancel,gensymb}
\definecolor{darkred}{rgb}{0.6,0,0} 
\definecolor{mightnightblue}{RGB}{25,25,112}
\definecolor{brown}{rgb}{0.59, 0.29, 0.0}
\hypersetup{colorlinks,citecolor= nicered,linkcolor= nicered} 

\definecolor{nicered}{rgb}{0.7,0.1,0.1} 
\definecolor{nicegreen}{rgb}{0.1,0.5,0.1}

\def \znbb {$\rm 0\nu\beta\beta$ }
\newcommand{\mli}{\ensuremath{m_{\rm lightest}}}
\newcommand{\mnu}{\ensuremath{\Sigma m_\nu}}

\newcommand{\Uaj}[1]{\ensuremath{|U_{#1}|^2}}
\newcommand{\ex}[1]{\ensuremath{\times10^{#1}}}
\newcommand{\Thl}{\ensuremath{T^{0\nu}_{1/2}}}
\newcommand{\Gps}{\ensuremath{G_{0\nu}^{\mathcal{N}}}}
\newcommand{\Mme}{\ensuremath{\mathcal{M}_{0\nu}^{\mathcal{N}}}}

\def\Valencia{Instituto de F\'{i}sica Corpuscular, CSIC-Universitat de Val\`{e}ncia, 46980 Paterna, Spain}

\AtBeginDocument{\hypersetup{citecolor=blue,linkcolor=nicered,urlcolor=Green3}}

\begin{document}

\title{\Large 2020 Global reassessment of the neutrino oscillation picture}

\author{P. F. de Salas}\email{pablo.fernandez@fysik.su.se}
\affiliation{The Oskar Klein Centre for Cosmoparticle Physics, Department of Physics, Stockholm University, AlbaNova, 10691 Stockholm, Sweden}
\author{D. V. Forero}\email{dvanegas@udem.edu.co}
\affiliation{Universidad de Medellín, Carrera 87 $N^o$ 30 - 65, Medellín, Colombia}
\author{S. Gariazzo}\email{gariazzo@to.infn.it}
\affiliation{\Valencia}
\affiliation{INFN, Sezione di Torino, Via P. Giuria 1, I--10125 Torino, Italy}
\author{P. Mart\'{i}nez-Mirav\'e}\email{pamarmi@ific.uv.es}
\affiliation{\Valencia}
\affiliation{Departament de Física Teòrica, Universitat de València, 46100 Burjassot, Spain}
\author{O. Mena}\email{omena@ific.uv.es}
\affiliation{\Valencia}
\author{C. A. Ternes}\email{chternes@ific.uv.es}
\affiliation{\Valencia}
\affiliation{INFN, Sezione di Torino, Via P. Giuria 1, I--10125 Torino, Italy}
\author{M. T{\'o}rtola}\email{mariam@ific.uv.es}
\affiliation{\Valencia}
\affiliation{Departament de Física Teòrica, Universitat de València, 46100 Burjassot, Spain}
\author{J. W. F. Valle}\email{valle@ific.uv.es}
\affiliation{\Valencia}

\begin{abstract}
\vskip .4cm
We present an updated global fit of neutrino oscillation data in the simplest three-neutrino framework.
In the present study we include up-to-date analyses from a number of experiments.
Concerning the atmospheric and solar sectors, besides the data considered previously, we give updated analyses of IceCube DeepCore and   Sudbury Neutrino Observatory data, respectively.
We have also included the latest electron antineutrino data collected by the Daya Bay and RENO reactor experiments, and the long-baseline T2K and NO$\nu$A measurements,  as reported in the Neutrino 2020 conference.
All in all, these new analyses result in more accurate measurements of $\theta_{13}$, $\theta_{12}$, $\Delta m_{21}^2$ and $|\Delta m_{31}^2|$.
The best fit value for the atmospheric angle $\theta_{23}$ lies in the second octant, but first octant solutions remain allowed at $\sim2.4\sigma$.
Regarding CP violation measurements, the preferred value of $\delta$ we obtain is 1.08$\pi$ (1.58$\pi$) for normal (inverted) neutrino mass ordering.
The global analysis still prefers normal neutrino mass ordering with 2.5$\sigma$ statistical significance. This preference is milder than the one found in previous global analyses.
These new results should be regarded as robust due to the agreement found between our Bayesian and frequentist approaches.
Taking into account only oscillation data, there is a weak/moderate preference for the normal neutrino mass ordering of $2.00
\sigma$.
While adding neutrinoless double beta decay from the latest Gerda, CUORE and KamLAND-Zen results barely modifies this picture, cosmological measurements raise the preference to $2.68\sigma$ within a conservative approach.
A more aggressive data set combination of cosmological observations leads to a similar preference for normal with respect to inverted mass ordering, namely $2.70\sigma$.
This very same cosmological data set provides $2\sigma$ upper limits on the total neutrino mass corresponding to $\mnu<0.12$ ($0.15$)~eV in the normal (inverted) neutrino mass ordering scenario.
The bounds on the neutrino mixing parameters and masses presented in
this up-to-date global fit analysis  include all currently available neutrino physics inputs.

\end{abstract}
\maketitle
\newpage\tableofcontents

\section{Introduction}
\label{sec:intro}
This paper updates the results from a long ongoing series of global fits to neutrino oscillation data~\cite{deSalas:2017kay,Forero:2014bxa,Tortola:2012te,Schwetz:2011zk,Schwetz:2011qt,Schwetz:2008er,Maltoni:2004ei,Maltoni:2003da}\footnote{For the results obtained by other groups see Refs.~\cite{Esteban:2020cvm,Capozzi:2020qhw}.}.
Neutrino flavor conversion was first observed in solar~\cite{Ahmad:2002jz} and atmospheric neutrinos~\cite{Fukuda:1998mi}.
This discovery led to the Nobel prize in Physics in 2015~\cite{McDonald:2016ixn,Kajita:2016cak} and was confirmed by subsequent results from the KamLAND reactor experiment~\cite{Eguchi:2002dm} as well as long baseline accelerator experiments.
These were crucial to identify neutrino oscillations as \textit{the} explanation of the solar neutrino problem and the atmospheric neutrino anomaly\footnote{Other mechanisms, such as magnetic moments~\cite{Miranda:2000bi,Miranda:2001hv,Barranco:2002te} or non-standard interactions~\cite{GonzalezGarcia:1998hj,Guzzo:2001mi,Miranda:2004nb} could be present only at a sub-leading level~\cite{Maltoni:2004ei}, for recent analyses see, e.g.~\cite{Esteban:2018ppq,Dev:2019anc}.}.
In the simplest three-neutrino scenario, the probability for a neutrino to oscillate between flavors is described by six parameters,
$\Delta m_{21}^2$, $|\Delta m_{31}^2|$, $\theta_{12}$, $\theta_{13}$, $\theta_{23}$ and $\delta$.
In addition, there are two possible mass orderings (MO) for neutrinos, according to the positive or negative sign of $\Delta m_{31}^2$.
In the first case, we talk about normal ordering (NO), and in the latter, about inverted ordering (IO).
The parameters are measured by different types of experiments, i.e.\ in solar experiments (SOL), in atmospheric experiments (ATM), in the long-baseline reactor experiment KamLAND, in short-baseline\footnote{Here, we use the term short-baseline for baselines of the order of $1$~km. 
We will not discuss the searches for light sterile neutrinos. We refer
the interested reader to Refs.~\cite{Gariazzo:2015rra,Gariazzo:2017fdh,Dentler:2018sju,Dentler:2017tkw,Gariazzo:2018mwd,Diaz:2019fwt,Giunti:2019aiy,Boser:2019rta}.}
reactor experiments (REAC) and in long-baseline accelerator experiments (LBL).
Moreover, data from cosmological observations (COSMO) can constrain the absolute mass scale, giving an indirect contribution to the determination of the neutrino mass ordering. If neutrinos turn out to be Majorana particles, the non-observation of $0\nu\beta\beta$ would also provide complementary information on the absolute neutrino mass scale and disfavor inverted neutrino mass ordering.

In Tab.~\ref{tab:exps} we summarize the sensitivity of the various experiment types in probing each of the oscillation parameters.
Since many of the parameters are measured by several classes of experiments, a combined or global fit of all data will give more precise results than a measurement of a single experiment on its own. Performing such global analysis is precisely the purpose of this study.
The paper is structured as follows: in Sec.~\ref{sec:exp} we present the analysis of each class of experiments, focusing on
solar experiments and KamLAND, short-baseline reactor experiments, atmospheric experiments and, finally, long-baseline accelerator experiments.
Next, we show the results from our global fit to neutrino oscillation data, following a frequentist approach in Sec.~\ref{sec:glob}, and a
Bayesian approach in Sec.~\ref{sec:bayesian}.
In Sec.~\ref{sec:mass} we discuss the effects of the inclusion of non-oscillation data sets and present our final results on the neutrino mass ordering. Finally, we summarize all our results in Sec.~\ref{sec:conc}.

\begin{table}
\centering
\begin{tabular}{|l|c|c|}
\hline
\text{Parameter}&\text{Main contribution from}&\text{Other contributions from}
\\
\hline
$\Delta m_{21}^2$& KamLAND & SOL
\\
$|\Delta m_{31}^2|$& LBL+ATM+REAC& -
\\
$\theta_{12}$ & SOL & KamLAND
\\
$\theta_{23}$ & LBL+ATM & -
\\
$\theta_{13}$& REAC & (LBL+ATM) and (SOL+KamLAND)
\\
$\delta$ & LBL & ATM
\\
MO & (LBL+REAC) and ATM & COSMO and $0\nu\beta\beta$
\\\hline
\end{tabular}
\caption{The main contribution to each of the oscillation parameters from the different classes of experiments.}
\label{tab:exps}
\end{table}

\section{Experimental data}
\label{sec:exp}

In this section we discuss the experimental results included in our global fit with more detail.
We dedicate one subsection to describe each class of experiments, discussing the main details of the data sets analyzed.
The results of the oscillation analysis in each sector are presented as well.

\subsection{Solar neutrino experiments and KamLAND}
\label{sec:exp_sol}

\begin{figure}[t!]
\centering
\includegraphics[width=0.5\textwidth]{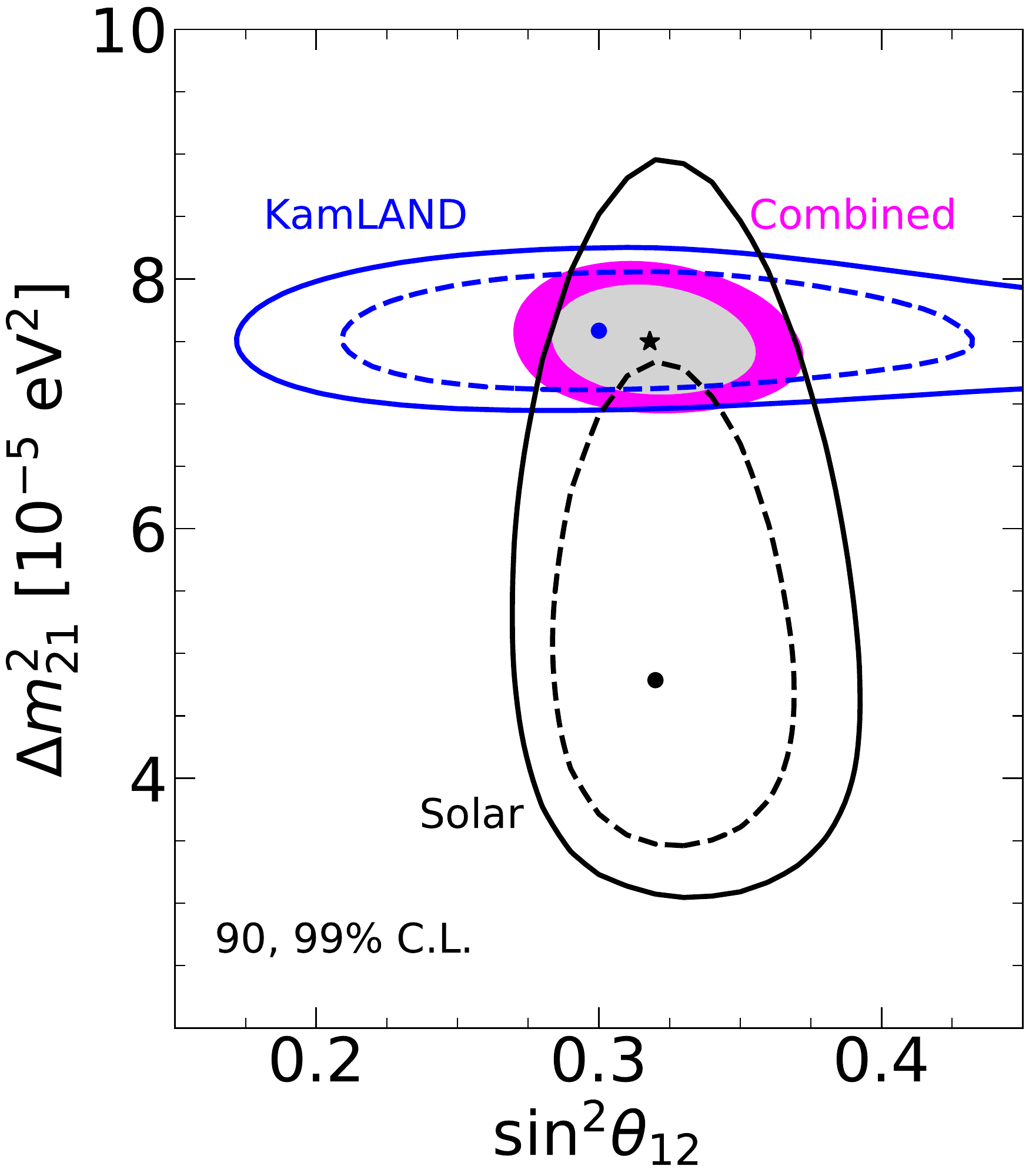}
\caption{90 and 99\% C.L.\ (2 d.o.f.) allowed regions in the
$\sin^2\theta_{12}$--$\Delta m^2_{21}$ plane obtained from the
analysis of solar neutrino experiments (black lines), from KamLAND
(blue lines) and from the combined analysis (colored regions). The
best fit values are indicated with dots for the independent analyses and with a star for the combined solar + KamLAND analysis. The reactor mixing angle $\theta_{13}$ has been marginalized over without further constraint from short-baseline reactor experiments.}
\label{fig:sol_exps}
\end{figure}

Solar neutrinos are produced in thermonuclear reactions in the interior of the Sun when burning hydrogen into helium. 
The main nuclear chains producing neutrinos are the so-called proton-proton ($pp$) chain and the CNO cycle. 
Neutrinos are produced in different reactions with energies ranging
from 0.1 to 20 MeV.
Our solar oscillation analysis includes data from all past and present solar neutrino oscillation experiments. 
We use the total rate measurements performed at the radiochemical experiments Homestake~\cite{Cleveland:1998nv}, 
GALLEX/GNO~\cite{Kaether:2010ag} and SAGE~\cite{Abdurashitov:2009tn},
the low-energy $^7$Be neutrino data from Borexino~\cite{Bellini:2011rx,Bellini:2013lnn},
as well as the zenith-angle or day/night spectrum from phases I--IV in Super-Kamiokande~\cite{Hosaka:2005um,Cravens:2008aa,Abe:2010hy,Nakano:PhD}. 
\footnote{
The measurement of the CNO solar neutrino flux, recently presented by Borexino \cite{gioacchino_ranucci_2020_4134014} is not expected to have an impact on the determination of the oscillation parameters and, hence, is not included in this analysis. Regarding the latest solar results from Super-Kamiokande~\cite{yasuhiro_nakajima_2020_4134680}, although they are also not included in this analysis, their implications are discussed.
}
Finally, we also include the last results from the Sudbury Neutrino Observatory (SNO), combining the solar neutrino data from the three phases of the experiment~\cite{Aharmim:2011vm}.
As in previous works, we have considered the low metallicity version of the standard solar model, labeled as AGSS09~\cite{Vinyoles:2016djt}.
The result of our combined analysis of solar neutrino oscillation data is shown in Fig.~\ref{fig:sol_exps}.

The solar neutrino oscillation parameters were also measured at the KamLAND experiment~\cite{Abe:2008aa,Gando:2010aa,Gando:2013nba}.
This long-baseline reactor neutrino experiment used a single detector to detect neutrinos from 56 nuclear reactors at an average distance of 180~km. 
This long distance made KamLAND sensitive to the values of the mass splitting $\Delta m^2_{21}$ indicated by the solar data analysis.
In our global fit, we include KamLAND data as presented in Ref.~\cite{Gando:2010aa}. 
The result of our analysis is shown together with the result from the analysis of solar neutrino oscillation data in Fig.~\ref{fig:sol_exps}.
As can be seen in the figure, the solar experiments provide a more precise measurement of the solar mixing angle, while KamLAND gives a better determination of the solar mass splitting.
Note that, since KamLAND is mostly sensitive to $\sin^22\theta_{12}$, using KamLAND data alone, we would obtain a second minimum in the upper octant of $\sin^2\theta_{12}$.
This solution is excluded when combining with solar neutrino data, sensitive to $\sin^2\theta_{12}$ through the observation of the adiabatic conversion in the solar medium. Note, however, that the upper-octant solution may emerge in the presence of non-standard interactions~\cite{Miranda:2004nb,Escrihuela:2009up,Coloma:2016gei}.

Recently, the final results from Super-Kamiokande IV were presented, including 2970 days of data taking \cite{yasuhiro_nakajima_2020_4134680}.
These results are particularly important due to two factors.
First of all, the ratio between the data and the unoscillated prediction has shifted upwards with respect to previous results.
Secondly, a smaller value of the day-night asymmetry has been reported.
According to the internal analysis of the collaboration, these two facts contribute equally, in terms of $\Delta \chi^2 $, to shift the preferred value of $\Delta m^2_{21}$ to higher values.
As a result, the previous tension between solar experiments and KamLAND in the determination of this parameter is significantly reduced.
Despite their relevance for the consolidation of the neutrino oscillations picture, these results are not expected to have a large impact in global fits, since the preferred value of $\Delta m^2_{21}$ after the combination of all data sets is dominated
by KamLAND~\footnote{In addition, publicly available information does not allow one to precisely reproduce the results reported \cite{yasuhiro_nakajima_2020_4134680} and hence these constraints are not included in our analysis due to the impossibility of
accounting properly for these measurements.}.

The best fit value obtained by solar experiments, $\Delta m_{21}^2 = 4.8\times10^{-5}$~eV$^2$, is excluded by KamLAND with a high confidence level.
However, the regions overlap above 90\% C.L.
Moreover, notice that the solar experiments and KamLAND show also a marginal sensitivity to $\theta_{13}$, which can be enhanced at the combined analysis~\cite{Goswami:2004cn,Schwetz:2008er}.
In order to generate Fig.~\ref{fig:sol_exps}, we have marginalized over
$\theta_{13}$, without taking any constraint from short baseline
reactor data, which we discuss in the next subsection.

\subsection{Reactor neutrino experiments}
\label{sec:exp_reac}

Besides KamLAND, there are several other reactor neutrino oscillation experiments.
Here, we use data coming from the reactor experiments RENO~\cite{Bak:2018ydk} and Daya Bay~\cite{Adey:2018zwh}.
Unlike KamLAND, they lie quite close to the nuclear power plants. This makes them sensitive to $\theta_{13}$ and 
$\Delta m_{31}^2$\footnote{Actually, short-baseline reactor experiments are sensitive to the effective mass splitting $\Delta m^2_{ee} = \cos^2\theta_{12}\Delta m_{31}^2 + \sin^2\theta_{12}\Delta m_{32}^2$~\cite{Nunokawa:2005nx}.}. 
Using current reactor neutrino data, it was shown that there is also some sensitivity to the solar parameters~\cite{Hernandez-Cabezudo:2019qko}.
Note that these, however, are not competitive with the results coming from KamLAND and solar experiments and therefore we fix in our
analyses the solar parameters to the ones measured by those experiments, as discussed in the previous section.

The Reactor Experiment for Neutrino Oscillation (RENO) is a neutrino oscillation experiment located at the Hanbit Nuclear Power Plant (South Korea), that has been taking data since August 2011. 
Two functionally identical $16$~ton detectors placed at $294$~m and $1383$~m from the 
centerline of the antineutrino sources, detect electron antineutrinos produced by six 
pressurized water reactors (all equally distributed in space along a $3$~km line), each 
with output thermal powers of $2.6~\text{GW}_{\text{th}}$ or $2.8~\text{GW}_{\text{th}}$. 
The average relative fission fractions for these reactor cores can be found in Ref.~\cite{Seo:2016uom}.
In the most recent publication~\cite{jonghee_yoo_2020_4123573}, the RENO collaboration reported results that correspond to $2900$ days of data taking,
updating their former findings~\cite{Bak:2018ydk}. 
From the observation of electron antineutrino disappearance, RENO reported a value for the reactor mixing angle of $\sin^2(2\theta_{13})=0.0892\pm 0.0063$, and a value of $|\Delta 
m^2_{ee}|=(2.74\pm0.12)\times10^{-3}\,\text{eV}^2$ for the observed neutrino mass squared 
difference. 
In our analysis, we consider antineutrino events (background subtracted) at the near and 
far detectors, as reported by RENO~\cite{jonghee_yoo_2020_4123573}, distributed along 26 energy bins 
in prompt energy, ranging from $1.2$~MeV to $8.0$~MeV. A total of nine systematical 
uncertainties, accounting for reactor-flux uncertainties $\sigma_r=0.9\%$ (correlated 
between detectors), uncorrelated detection uncertainty 
$\sigma_{du}=0.21\%$~\cite{RENO:2015ksa,Seo:2016uom}, and an overall normalization 
uncertainty $\sigma_o=2\%$, have been included in the analysis.
In the calculation of the signal events, a Gaussian energy smearing was assumed to account for the detector energy 
resolution with a width $\sigma_E/E\approx7\%/\sqrt{E[\text{MeV}]}$~\cite{Seo:2016uom}.

The Daya Bay Reactor Neutrino experiment analyzes the antineutrino flux produced by six reactor cores at the Daya Bay and Ling Ao nuclear power plants. 
The electron antineutrino oscillation probability is measured by eight identical antineutrino detectors (ADs).
Two detectors are placed in each of the two near experimental halls of the experiment (EH1 and EH2), while the remaining four are located at the far experimental hall (EH3).
Detailed studies on the antineutrino flux and spectra have been performed in order to determine the fission fractions (see Tab.~9 in Ref.~\cite{An:2016srz}) as well as the thermal power (see Tab.~I in Ref.~\cite{An:2016ses}).
Baseline distances range in $\sim$0.3 -- 1.3~km for the near experimental halls and $\sim$1.5 -- 1.9~km for the far hall.
The Daya Bay collaboration analyzed data collected after 1958 days of running time~\cite{Adey:2018zwh} and reported the measurements $\sin^2(2\theta_{13})=0.0856\pm 0.0029$ and $|\Delta m^2_{ee}|=(2.522^{+0.068}_{-0.070})\times10^{-3}\,\text{eV}^2$.
To obtain the oscillation parameters, our analysis uses the number of antineutrino events after background subtraction, considering the ratios of EH3 to EH1 and EH2 to EH1.
Regarding the statistical methods, the Daya Bay collaboration has followed three different approaches (covariant approach, nuisance parameters and a hybrid approach).
Consistent results can be obtained with the three methods and we have chosen to use nuisance parameters in our analysis.
The uncertainties arising from the power and fission fractions at each
of the 6 nuclear reactors are encoded in these nuisance parameters
($\sigma_r = 0.2\%$ and $\sigma_{frac}= 0.1\%$). In addition,
characteristics of each detector, such as the differences in the
running time or the efficiencies, have been accounted for in the simulation.
Other sources of uncertainties, such as shifts in the energy scale
($\sigma_{scale} = 0.6\%$), have also been included.

\begin{figure}[t!]
\centering
\includegraphics[width=0.85\textwidth]{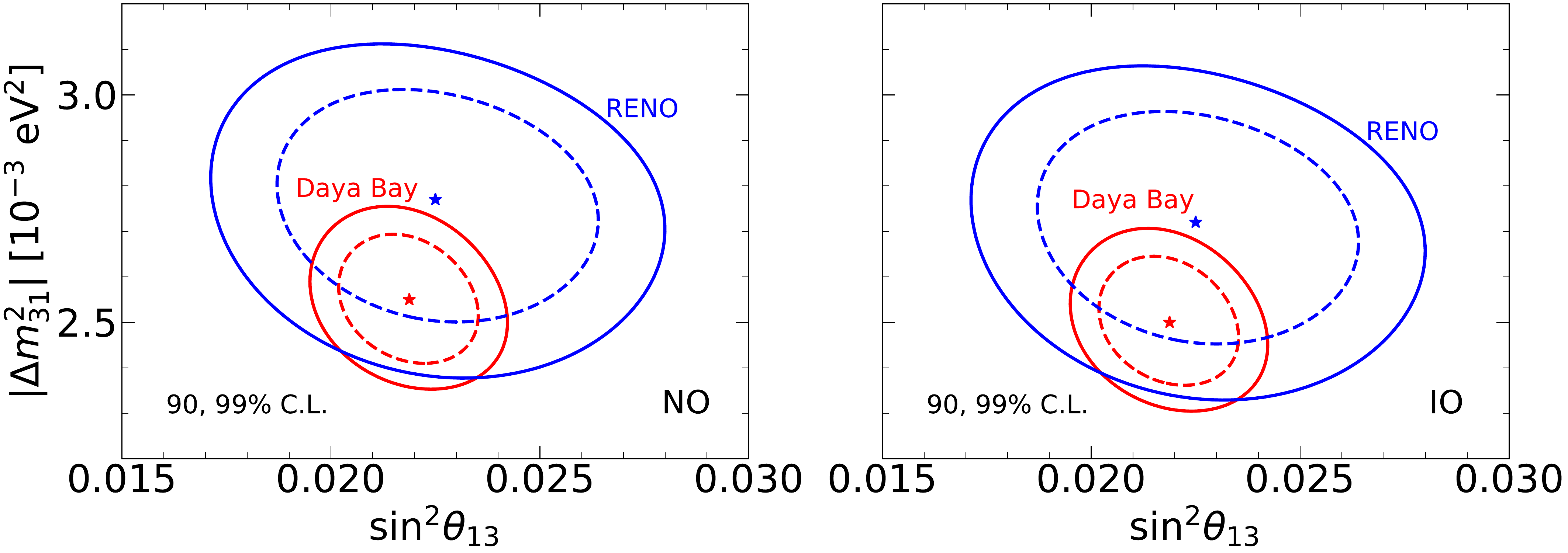}
\caption{90 and 99\% C.L.\ (2 d.o.f.) allowed regions in the
$\sin^2\theta_{13}$--$\Delta m^2_{31}$ plane for RENO (blue) and
Daya Bay (red). The best fit values are indicated by stars. The
left (right) panels correspond to NO (IO).}
\label{fig:reac_exps}
\end{figure}

The results of our analyses of short-baseline reactor data are shown in Fig.~\ref{fig:reac_exps}.
As can be seen in the figure, there is a total overlap between the parameter regions determined by RENO and Daya Bay, although the latter clearly dominates the measurement of the relevant oscillation parameters. 
Note also that our results are almost identical for normal and inverted mass spectra, since these experiments are not sensitive to the mass ordering.

\subsection{Atmospheric neutrino experiments}
\label{sec:exp_atm}

When cosmic rays collide with particles in the Earth's atmosphere, they start a particle shower which eventually creates the atmospheric neutrino flux.
The energy of $\nu_\mu$ and $\nu_e$ (and their antiparticles) produced in the atmosphere can range from a few MeV up to roughly $10^9$~GeV, although only events up to $\sim100$~TeV are currently detectable.
The energy of the atmospheric neutrinos relevant to oscillation studies, however, ranges from $\sim0.1$~GeV to $\sim100$~GeV.
In our global fit we include data
from Super-Kamiokande~\cite{Abe:2017aap} and from IceCube DeepCore~\cite{Aartsen:2017nmd,Aartsen:2019tjl}.
Since the largest part of the atmospheric neutrino flux is composed by $\nu_\mu$ and $\overline{\nu}_\mu$, and given that it is more difficult to identify electrons in the detector, the main channel used in current atmospheric neutrino experiments is $\nu_\mu\to\nu_\mu$, which makes them mostly sensitive to the oscillation parameters $\theta_{23}$ and $\Delta m_{31}^2$.
Note, however, that the Super-Kamiokande experiment also detected a large sample of electron events from $\nu_e$ appearance~\cite{Abe:2017aap,Jiang:2019xwn}.
The analysis  of these results, however, can not be precisely performed outside the experimental collaboration.
As a result, we do not analyze Super-Kamiokande atmospheric data ourselves, but only include the latest $\chi^2$-table made available by the collaboration~\cite{SKIV-tabs}.
The Super-Kamiokande collaboration recently presented an updated analysis of atmospheric neutrino data~\cite{yasuhiro_nakajima_2020_4134680}.
This new analysis slightly prefers the region with $\sin^2\theta_{23}<0.5$ and shows a weaker preference for the normal neutrino mass ordering than the data set included in our analysis. However, the collaboration has not released the new $\chi^2$-tables and, therefore,  we can not include the new atmospheric results in our present analysis. 

\begin{figure}
\centering
\includegraphics[width=0.85\textwidth]{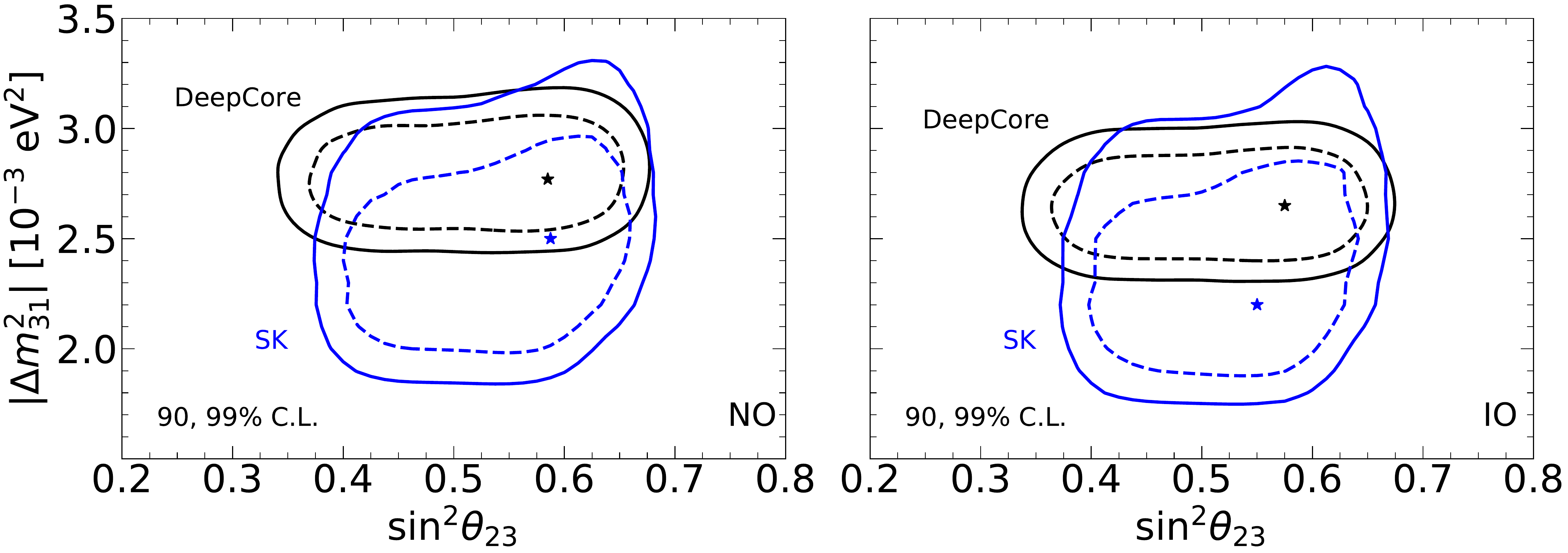}
\caption{
90 and 99\% C.L.\ (2 d.o.f.) allowed regions at the
$\sin^2\theta_{23}$--$\Delta m^2_{31}$ plane for NO (left) and
IO (right), obtained from the analyses of
Super-Kamiokande (SK) atmospheric (blue) and DeepCore (black) data. The
best fit values are indicated by stars.}
\label{fig:atm_exps}
\end{figure}

For the current global fit, we update our analysis of DeepCore data. In addition to track-like events, the released experimental data now includes also shower-like events, increasing the number of events from roughly 6000~\cite{Aartsen:2014yll} to around 20000~\cite{Aartsen:2017nmd,Aartsen:2019tjl}.
The data analyzed correspond to 3 years of observations of the full sky, from April 2012 to May 2015.
The details of the analysis are described in Ref.~\cite{Aartsen:2019tjl}, and the full data set can be downloaded from Ref.~\cite{SampleA:IceCube-August-2019}.
Two data samples are provided: Sample A and Sample B, corresponding to the same data taking period but different cuts.
For this analysis we have chosen Sample A.
Several sources of systematic uncertainties are included in our analysis. They can be divided into detector-related and flux-related uncertainties. 
We account for neutrino scattering and absorption in the ice, and include several uncertainties related to the optical efficiencies.
Concerning the atmospheric neutrino flux, we include systematic uncertainties on the ratio of neutrinos to antineutrinos, the ratio of electron to muon neutrinos, the spectral index, the ratio of vertically to horizontally incoming neutrinos and an overall normalization.
The results of our analysis are depicted in Fig.~\ref{fig:atm_exps}, together with the ones from Super-Kamiokande.
As in the reactor case, the regions allowed by the two experiments totally overlap.
However, one can see that the mixing angle is slightly better measured by Super-Kamiokande, while DeepCore provides a more stringent result on the atmospheric mass splitting.

\subsection{Accelerator experiments}
\label{sec:exp_lbl}

\begin{figure}
\centering
\includegraphics[width=0.85\textwidth]{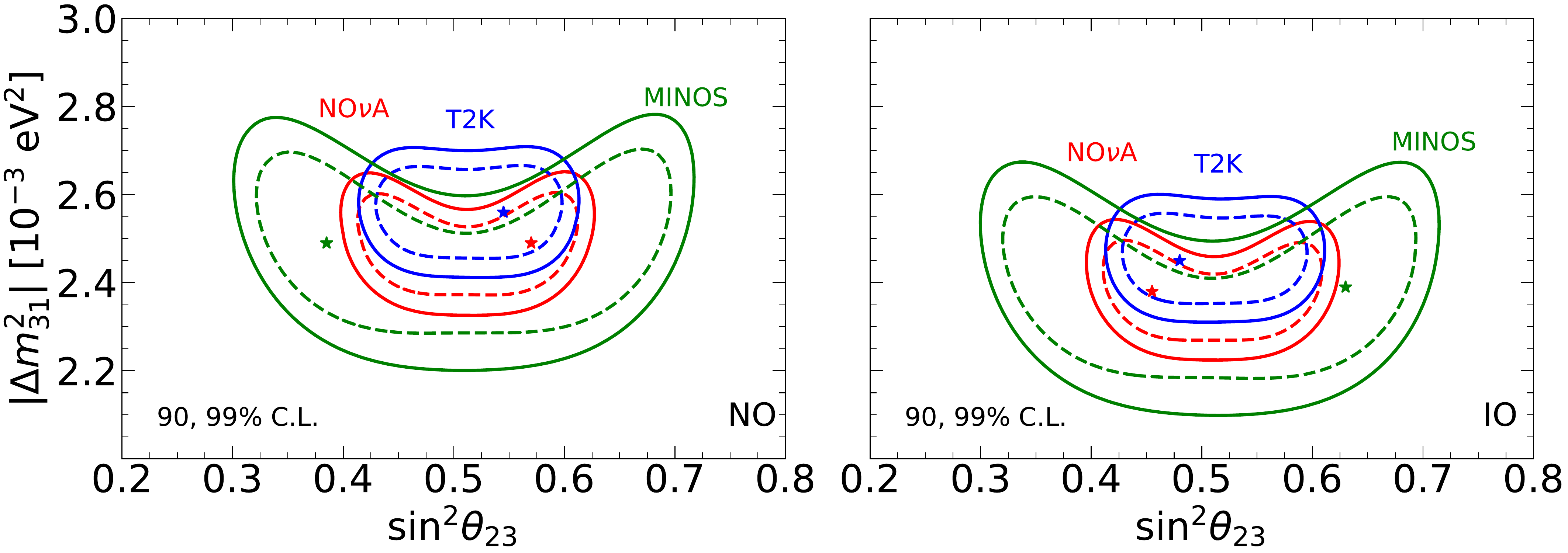}
\caption{
90 and 99\% C.L.\ (2 d.o.f.) allowed regions in the
$\sin^2\theta_{23}$--$\Delta m^2_{31}$ plane for NO (left) and IO (right), obtained from the analyses of T2K (blue), NO$\nu$A (red) and MINOS (green) data. The best fit values are
indicated by stars.}
\label{fig:lbl_exps1}
\end{figure}
\begin{figure}
\centering
\includegraphics[width=0.85\textwidth]{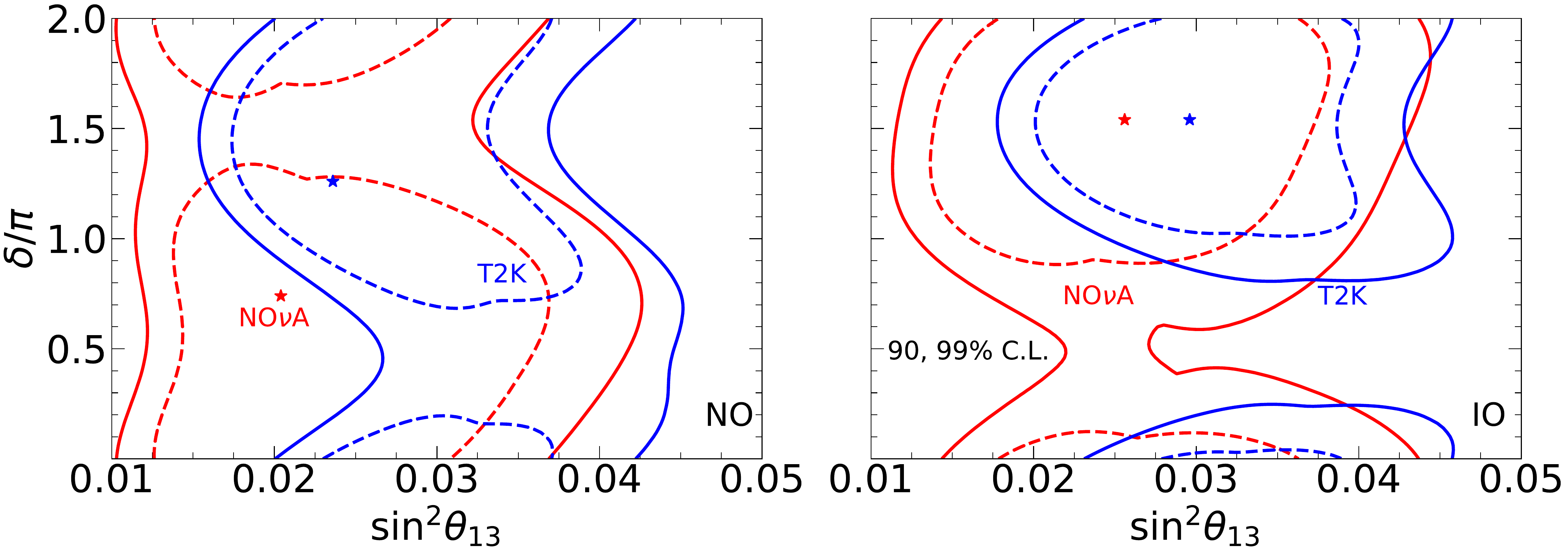}
\caption{
90 and 99\% C.L.\ (2 d.o.f.) allowed regions in the $\sin^2\theta_{13}$--$\delta$ plane for NO (left) and IO (right), obtained from the analyses of T2K (blue) and NO$\nu$A (red) data. The best fit values are indicated by stars.}
\label{fig:lbl_exps2}
\end{figure}

Long-baseline accelerator neutrino experiments measure neutrinos which are created in particle accelerators.
They originate in meson decays. The mesons, typically pions and kaons, are created in the accelerator and then focused into a beam.
Next, they decay into muon-neutrinos, while a beam dump absorbs the ones which do not decay.
Using different polarities of the focusing horns one can separate mesons from antimesons, resulting in a mostly pure beam of neutrinos or antineutrinos.
Note, however, that creating a really pure beam is not possible, and there will always be a background contamination of so-called ``wrong-sign'' neutrinos.
The long-baseline experiments consist of two detectors, one near detector measuring the initial neutrino flux, close to the accelerator complex, and a far detector measuring the oscillated neutrino flux.
Long-baseline experiments measure the appearance of $\nu_e$ from the initial $\nu_\mu$ flux, and also the disappearance of $\nu_\mu$. This makes them sensitive to the oscillation parameters $\Delta m_{31}^2$, $\theta_{23}$, $\theta_{13}$, $\delta$ and, in principle, also to the neutrino mass ordering.
In our global fit, we use data from several long-baseline experiments: NO$\nu$A~\cite{alex_himmel_2020_3959581}, T2K~\cite{patrick_dunne_2020_3959558}, MINOS~\cite{Adamson:2014vgd} and K2K~\cite{Ahn:2006zza}.

The T2K collaboration has presented an updated analysis of neutrino and antineutrino data, corresponding to an exposure at Super-Kamiokande of 1.97$\times10^{21}$ protons on target (POT) in neutrino mode and 1.63$\times10^{21}$ POT in antineutrino mode. Data have been collected from January 2010 until June 2018.
T2K observed 318 (137) muon (anti-muon) events and 94 (16) electron (positron) events.
In addition, 14 electron events where also a pion is produced were recorded.
These results improve their former ones~\cite{Abe:2019ffx,Abe:2019vii,Abe:2018wpn}, allowing them now to exclude CP-conserving values of $\delta$ at close to 3$\sigma$ confidence level.

On the other hand, NO$\nu$A has reached 13.6$\times10^{20}$~POT in neutrino mode~\cite{NOvA:2018gge} and 12.5$\times10^{20}$~POT in antineutrino mode. NO$\nu$A finds 212 (105) muon (anti-muon) events and 82 (33) electron (positron) events.
The events in antineutrino mode constitute the first ever significant observation of $\overline{\nu}_e$ appearance in a long baseline experiment~\cite{Acero:2019ksn}.
Unlike T2K, the latest neutrino and antineutrino NO$\nu$A data prefer values of the CP-violating phase $\delta$ close to 0.8$\pi$ for normal ordering,
in tension with the T2K result.

In order to perform our analysis, we extract the relevant data for each experiment from the corresponding reference. We simulate the signal and background rates using the GLoBES software~\cite{Huber:2004ka,Huber:2007ji}.
For the energy reconstruction we assume Gaussian smearing. We include bin-to-bin efficiencies, which are adjusted to reproduce the best-fit spectra reported in the corresponding references.
Finally, for our statistical analysis we include systematic uncertainties, related to the signal and background predictions, which we minimize over.
The results of our analysis (without a prior on $\theta_{13}$) are presented in Figs.~\ref{fig:lbl_exps1} and~\ref{fig:lbl_exps2}. 
We find that T2K and NO$\nu$A measure the atmospheric parameters $\theta_{23}$ and $|\Delta m^2_{31}|$ rather well and with similar sensitivity.
Note, however, that T2K shows a slightly better sensitivity to $\theta_{13}$ and $\delta$ for inverted neutrino mass ordering, as
    indicated by the 90\% C.L.\ closed regions in the right panel of Fig.~\ref{fig:lbl_exps2}.
For normal neutrino mass ordering, both experiments show similar sensitivity to $\delta$, although T2K provides a better measurement of
the mixing angle $\theta_{13}$ than NO$\nu$A. 
In any case, these results are not competitive with short-baseline reactor experiments, discussed above. 
Focusing on the determination of $\delta$, the aforementioned tension between T2K and NO$\nu$A results for normal ordering is clearly visible in the left panel of Fig.~\ref{fig:lbl_exps2}. Note that here we are not imposing any prior on $\theta_{13}$, as the experimental collaborations do, and yet, the mismatch between both samples is quite evident. 
We shall discuss in more detail this tension in the measurement of $\delta$ and its consequences on the determination of the neutrino mass ordering in Sec.~\ref{sec:glob}.

We also show the results from our analysis of MINOS data~\cite{Adamson:2013whj,Adamson:2013ue}, which still contributes to the determination of $|\Delta m_{31}^2|$, as seen in Fig.~\ref{fig:lbl_exps1}.
Unfortunately, in this case there is no sensitivity to $\theta_{13}$ and $\delta$.
The same applies to the pioneering K2K experiment~\cite{Aliu:2004sq}, included in our global fit as well, but with a sensitivity to the oscillation parameters which has been overcome by the more recent long-baseline accelerator experiments.

\section{Results from the global fit}
\label{sec:glob}

\begin{figure}
\centering
\includegraphics[width=0.85\textwidth]{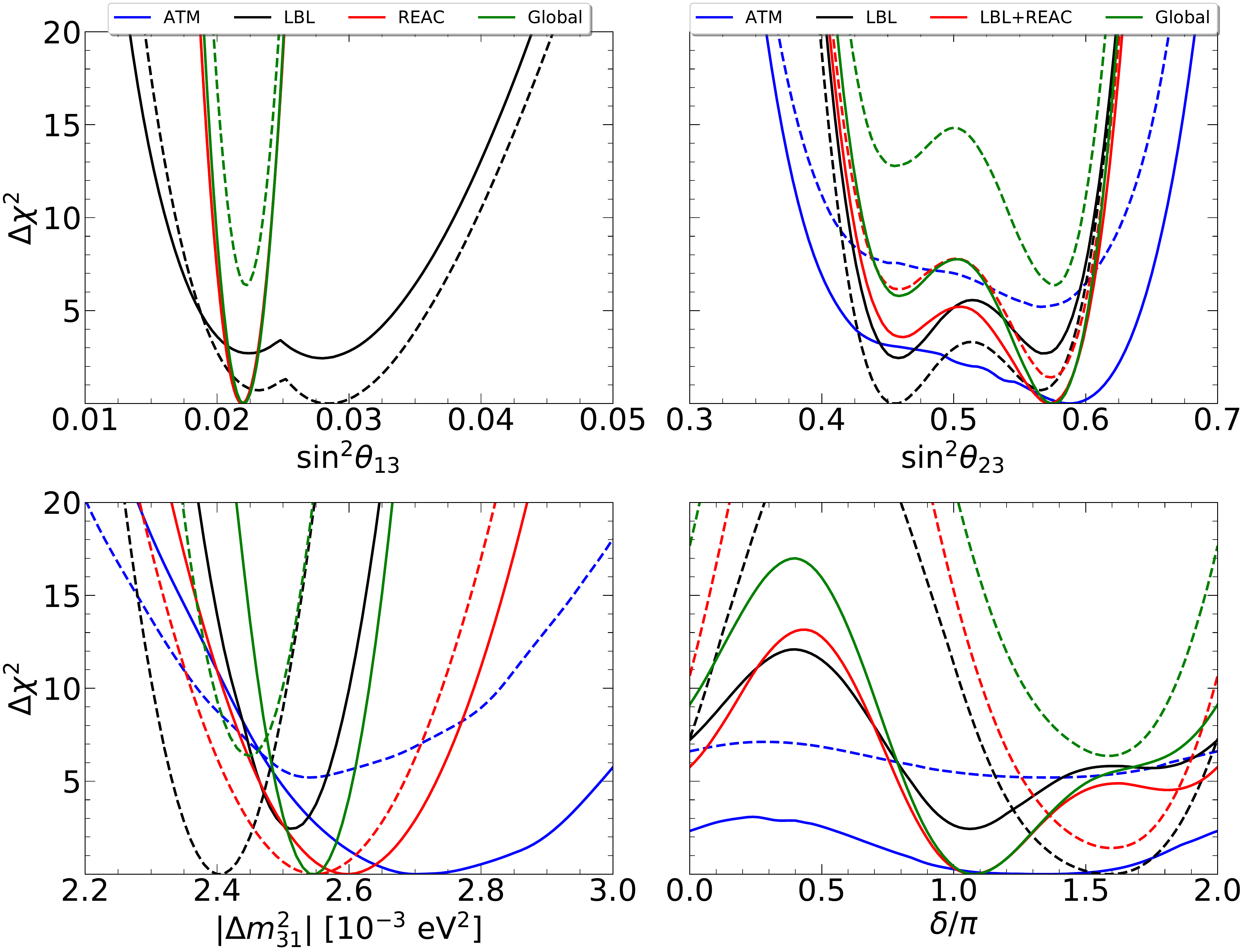}
\caption{
$\Delta\chi^2$ profiles obtained from the combination of all ATM data (blue), all LBL data (black) and global data (green).
Red lines correspond to the analysis of all REAC data in the panels of the parameters measured by reactor experiments directly and to LBL+REAC data for the parameters where reactors enter only via correlations. The solid (dashed) lines correspond to NO (IO).
The profiles are calculated with respect to the global minimum for each data sample, corresponding to normal ordering in all cases.}
\label{fig:chi2_profiles_exps}
\end{figure}

In the previous section, we have presented the individual results of our neutrino data analysis, obtained sector-by-sector.
In this section, we shall describe the results obtained by combining all previous data into our global neutrino oscillation fit.
We will first briefly discuss the main contributions to the well-measured parameters, and then enter into more detail in the discussion of the remaining unknowns of the three-neutrino picture.

\subsection{Well-measured oscillation parameters}
\label{sec:precise}

So far the solar parameters $\theta_{12}$ and $\Delta m_{21}^2$ have only
been measured by KamLAND and the solar neutrino experiments, and they
have already been discussed in Sec.~\ref{sec:exp_sol}.
After combining
with data from other experiments the determination of the solar
parameters improves further, due to a better determination of $\theta_{13}$,
but the effect is not very visible.
The future reactor experiment JUNO is expected
to measure the solar parameters with great precision~\cite{An:2015jdp}.
In contrast, the measurement of the remaining oscillation
parameters emerge from the combinations of several data sets, as seen in Fig.~\ref{fig:chi2_profiles_exps}.
From these four parameters, only $\theta_{13}$ and $\Delta m_{31}^2$ have been already measured with good precision at oscillation experiments.
Concerning the reactor mixing angle, if we compare the regions in
Fig.~\ref{fig:reac_exps} with those in Fig.~\ref{fig:lbl_exps2}, one sees that
the measurement of $\theta_{13}$ is clearly dominated by reactor experiments.
The contribution from other experiments to this result is negligibly small.
This behavior can also be appreciated in the upper left panel of Fig.~\ref{fig:chi2_profiles_exps},
where we see that the combination of global data (green lines) is basically equivalent to the combination of reactor data (red lines).
Regarding the absolute value of the atmospheric mass splitting, $|\Delta m_{31}^2|$, we see from Figs.~\ref{fig:reac_exps},~\ref{fig:atm_exps} and~\ref{fig:lbl_exps1} that 
its determination comes mainly from long-baseline accelerators and from Daya Bay,
although the determination by atmospheric experiments is still important,
as indicated in the lower left panel of Fig.~\ref{fig:chi2_profiles_exps}.
Comparing the lines corresponding to the analyses of long-baseline (black), reactor (red) and atmospheric data (blue) together with the result from the global fit (green lines) we find that,
unlike the case of $\sin^2\theta_{13}$, all experiments contribute significantly to this measurement.
%

\subsection{The atmospheric angle $\theta_{23}$}
\label{sec:atm-octant}

\begin{figure}
\centering
\includegraphics[width=0.85\textwidth]{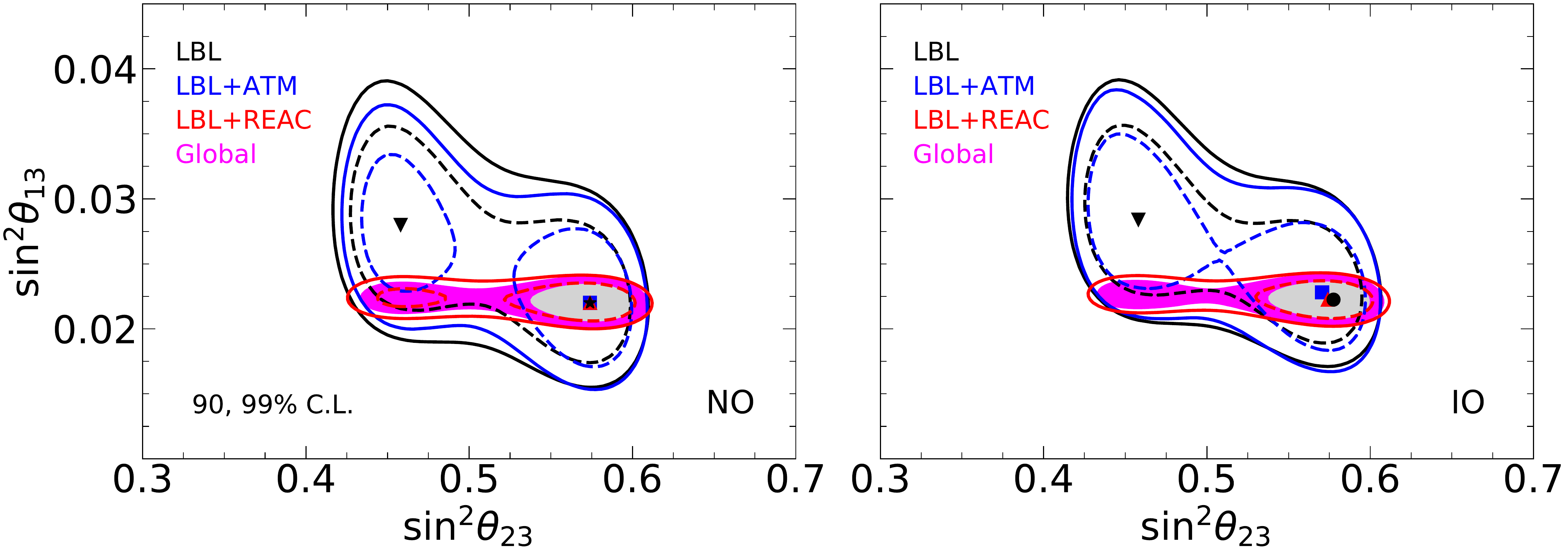}
\caption{Allowed regions in the $\sin^2\theta_{23}$--$\sin^2\theta_{13}$ plane from several combinations of data sets:
 LBL (black lines), LBL+ATM (blue), LBL+REAC (red) and global analysis (colored regions). The down-triangle, square and up-triangle correspond to the best fit values obtained from the combination of
 data sub-sets, while the black star (left panel) and black dot (right panel) denote the best value obtained from the global fit, for normal and inverted ordering, respectively.}
\label{fig:sq23_sq13_exps}
\end{figure}

Next, we discuss the determination of the atmospheric mixing angle, $\theta_{23}$.
Accelerator and atmospheric oscillation experiments measure the disappearance of muon (anti)neutrinos and are mainly sensitive to $\sin^22\theta_{23}$. 
Therefore, they can not resolve the octant of the angle: in other words, they can not determine if $\sin^2\theta_{23}>0.5$ or $\sin^2\theta_{23}<0.5$.
However, due to matter effects in the neutrino trajectories inside the Earth, this degeneracy is slightly broken for atmospheric neutrino oscillation experiments, see the blue lines in the upper right panel of Fig.~\ref{fig:chi2_profiles_exps}.
Also, the quantity $\sin^2\theta_{23}$ enters directly in the appearance channels of these experiments and, hence, the degeneracy can be further broken when including the electron neutrino samples in the fit.
Analyzing the data from long-baseline accelerators, we find two essentially degenerate solutions for $\sin^2\theta_{23}$ for both mass orderings, as indicated by the black lines in the upper right panel of Fig.~\ref{fig:chi2_profiles_exps}. 
The best fit is obtained for $\sin^2\theta_{23} = 0.46$, but a local minimum appears at $\sin^2\theta_{23} = 0.57$ with $\Delta\chi^2\approx 0.3$ (0.7) for normal (inverted) ordering.
Although $\theta_{23}$ is not measurable in reactor neutrino experiments, their data help in the determination of $\theta_{23}$ by breaking a degeneracy between $\theta_{23}$ and $\theta_{13}$, as indicated by the red lines in the upper right panel of Fig.~\ref{fig:chi2_profiles_exps}, obtained from the combination of reactor and long-baseline accelerator data.
In this case, the best fit value is obtained in the upper octant for both orderings.
This effect can be further appreciated in Figure~\ref{fig:sq23_sq13_exps}, showing the regions obtained from several combinations of data sets in the $\sin^2\theta_{23}$ - $\sin^2\theta_{13}$ plane at 90 and 99\% C.L.\ for two degrees of freedom. 
There, one sees how the combination of all accelerator data (black lines in the figure) prefers a rather large value of $\sin^2\theta_{13}$.
The combination of LBL with atmospheric data (blue) does not improve the determination of $\theta_{13}$, but shifts the best fit value of the analysis (indicated by the blue square) towards smaller values, as preferred by reactor data.
Note also that this combined analysis shifts the best fit value of $\sin^2\theta_{23}$ to the second octant.
A more distinctive feature appears when combining LBL with reactor data.
As expected, this combination results in a much more restricted range for $\theta_{13}$ and, therefore, the partial breaking of the
$\theta_{23}$-$\theta_{13}$ degeneracy, arising from the LBL appearance data, see the red lines in Fig.~\ref{fig:sq23_sq13_exps}.
Finally, when combining all data, we obtain the green lines in Fig.~\ref{fig:chi2_profiles_exps} and the colored regions in Fig.~\ref{fig:sq23_sq13_exps}. There, one sees that the effect from both combinations (LBL+ATM and LBL+REAC) is indeed very relevant for the determination of the octant of the atmospheric angle.
After combining all data samples, we obtain the best fit value of $\theta_{23}$ in the upper octant, with lower octant solutions
  slightly disfavored with $\Delta\chi^2 \geq 5.8$ (6.4) for normal (inverted) mass ordering. Maximal atmospheric mixing is disfavoured
  with $\Delta\chi^2 = 7.8$ (8.5) for normal (inverted) ordering\footnote{Note that the preference for the second octant, as well as the  rejection against maximal atmospheric mixing, could change when including the latest Super-Kamiokande results~\cite{yasuhiro_nakajima_2020_4134680}, not publicly available yet.}.

\subsection{The CP phase $\delta$}
\label{sec:delta}

\begin{figure}
\centering
\includegraphics[width=0.85\textwidth]{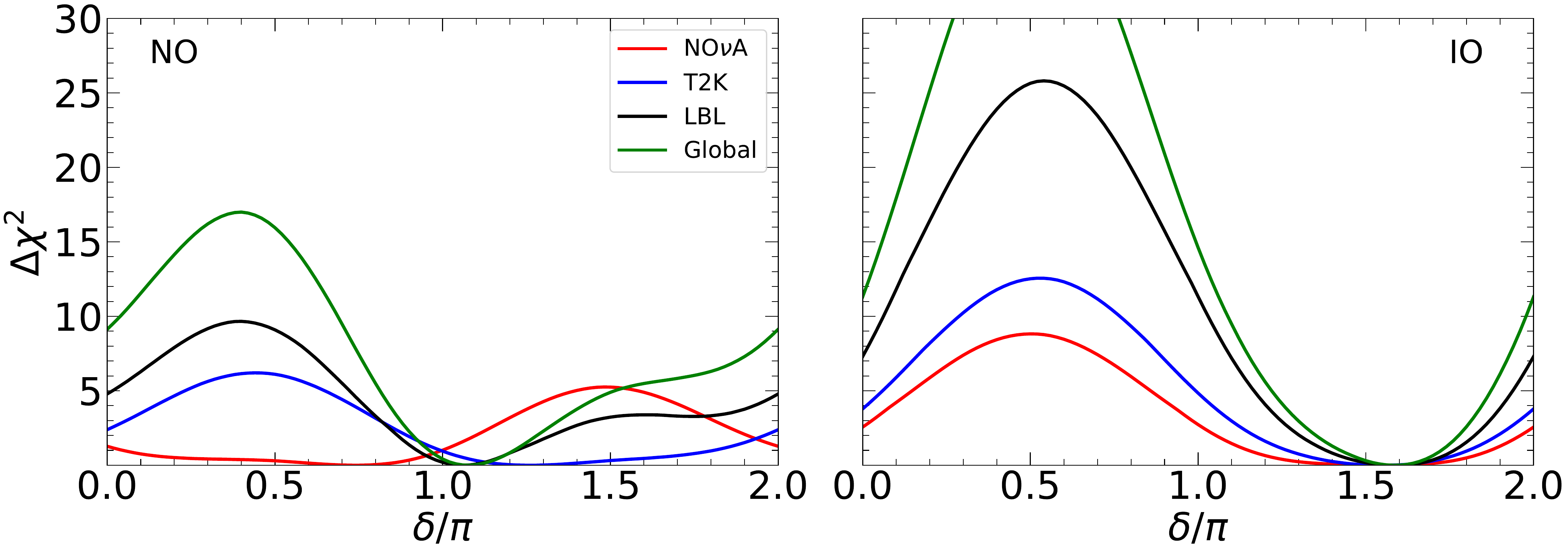}
\caption{$\Delta\chi^2$ profiles for $\delta$ obtained from the analysis of NO$\nu$A (red), T2K (blue), all long-baseline data (black) and from the global fit (green).}
\label{fig:delta_exps}
\end{figure}

\begin{figure}
\centering
\includegraphics[width=0.85\textwidth]{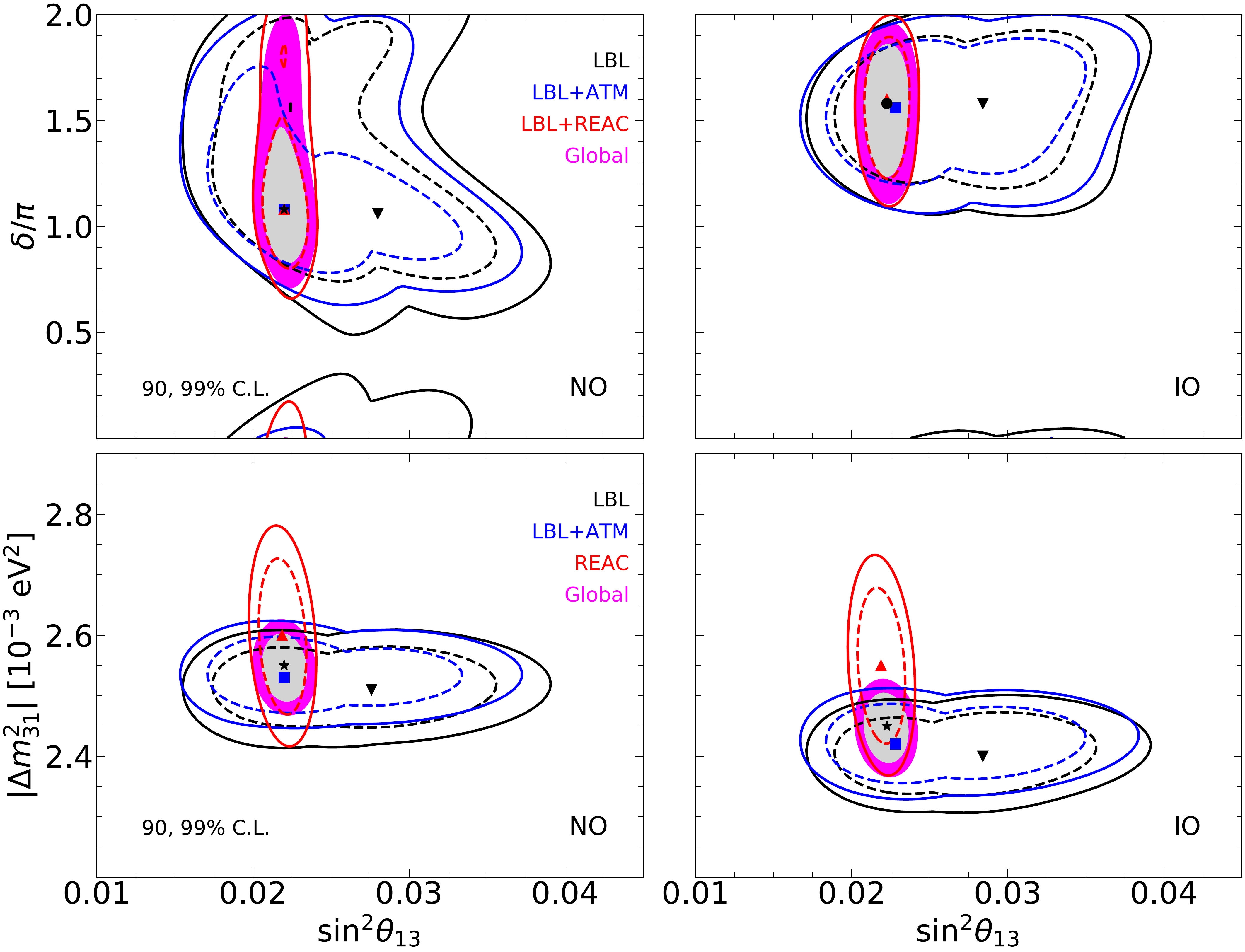}
\caption{Allowed regions in the $\sin^2\theta_{13}$-$\delta$ and $\sin^2\theta_{13}$-$\Delta m_{31}^2$ planes from several data-set combinations:
long-baseline accelerator experiments (black lines), accelerator and atmospheric data (blue), accelerator and reactor data (red, upper panels), reactor data (red, lower panels) and global analysis (colored regions). The down-triangle, square and up-triangle correspond to the best fit values obtained from the combination of data sub-sets, while the black star (left panel) and black dot (right panel) denote the best fit value obtained from the global fit for normal and inverted ordering, respectively.}
\label{fig:sq13_del_exps}
\end{figure}

We now discuss the measurement of the CP-violating phase, $\delta$. 
This phase induces opposite shifts in the $\nu_\mu\to\nu_e$ and $\overline{\nu}_\mu\to\overline{\nu}_e$ oscillation probabilities and, therefore, information on this parameter can be obtained by analyzing neutrino and antineutrino oscillation data in the appearance channels. 
Note, however, that the separate analysis of neutrino and antineutrino channels can not provide, at present, a sensitive measurement of $\delta$~\cite{Tortola:2020ncu}. 
The CP phase can therefore be measured by the long-baseline accelerator experiments T2K and NO$\nu$A, and also by Super-Kamiokande atmospheric neutrino data,
see the black and blue lines in the lower right panel of Fig.~\ref{fig:chi2_profiles_exps}. 
In addition to Fig.~\ref{fig:chi2_profiles_exps}, in Fig.~\ref{fig:delta_exps} we show the $\Delta\chi^2$ profiles for the CP-violating phase $\delta$ as obtained from the analysis of data from T2K (blue) and NO$\nu$A (red), the combination of all long-baseline data (black) and the result from the global fit (green). 
For normal neutrino mass ordering (left panel), a  tension arises between the determinations of $\delta$ obtained from T2K and NO$\nu$A data\footnote{This tension has
  been recently discussed in Refs.~\cite{Esteban:2020cvm} and~\cite{Kelly:2020fkv}.}. 
 Indeed, the analysis of NO$\nu$A results shows a preference for $\delta \approx 0.8\pi$, disfavoring the region around
  $\delta\approx1.5\pi$, where the best fit value for T2K is found.
This does not happen for inverted ordering (right panel), for which NO$\nu$A shows better sensitivity to $\delta$ and also an excellent agreement with T2K.
Note that this behavior is due to the antineutrino data sample collected by NO$\nu$A, and it is the reason why our sensitivity to $\delta$ in the current global fit is worse than it was in Ref.~\cite{deSalas:2017kay}.
The inclusion of reactor data can help to improve the determination of $\delta$, due to the existing correlation between the CP phase and $\theta_{13}$. 
This is illustrated in Fig.~\ref{fig:chi2_profiles_exps} and in the upper panels of Fig.~\ref{fig:sq13_del_exps}.
 From the global combination, we obtain the best fit value for the CP phase at $\delta = 1.08\pi$ (1.58$\pi$) for NO (IO). 
The CP-conserving value $\delta = 0$ is disfavored with $\Delta\chi^2 = 9.1$ (11.3). 
However, the other CP-conserving value, $\delta = \pi$, remains allowed with $\Delta\chi^2 = 0.4$ in NO, while it is excluded with $\Delta\chi^2 = 14.6$ in IO.

\subsection{The neutrino mass ordering}
\label{sec:nu-ord}

Finally, in this subsection, we present the results of our present analysis on the neutrino mass ordering issue. 
 Combining all neutrino oscillation data, we obtain a preference for normal mass ordering with respect to the inverted one with a
  value of $\Delta\chi^2 = 6.4$.
This corresponds to a 2.5$\sigma$ preference in favor of NO.
This preference comes from several contributions, which we shall discuss in the following. 
Our independent analyses of NO$\nu$A and T2K data do not show a particular preference for any mass ordering, since we obtain
$\Delta\chi^2 \approx 0.4$ in favor of NO in both cases.
Such a small value is expected, due to the rather small matter effects present in the neutrino propagation over the corresponding baselines. 
~However, after combining all the  long-baseline accelerator data, we find that IO is preferred with 2.4 units in $\Delta\chi^2$.
This result appears as a consequence of the tension in the measurement of $\delta$ by T2K and NO$\nu$A, as discussed in the previous subsection. 
Since the tension appears only in normal ordering, the minimum $\chi^2$ from the combined long-baseline analysis for this ordering is worse than the sum of the individual T2K and NO$\nu$A fits.
If we now perform a combined analysis of accelerator and reactor data, we obtain a preference for NO with $\Delta\chi^2 = 1.4$. 
The improved status of normal mass ordering comes from the difference in the measurements of $\Delta m^2_{31}$ in accelerator and reactor experiments.
As shown in Fig.~\ref{fig:chi2_profiles_exps} and in the lower panels of Fig.~\ref{fig:sq13_del_exps}, the values of $\Delta m^2_{31}$
preferred by accelerator and reactor experiments show a better agreement for normal ordering than for the inverted one.
On the other hand, the atmospheric neutrino results from the Super-Kamiokande and DeepCore experiments show some sensitivity to the neutrino mass ordering on their own. 
From Super-Kamiokande data alone (neither imposing a prior on $\theta_{13}$ nor combining with data from reactor experiments), there is already a preference for normal mass ordering with $\Delta\chi^2 \approx 3.5$, while DeepCore gives $\Delta\chi^2 \approx 1.0$. 
Combining the atmospheric neutrino results with long-baseline accelerator data, the preference for normal ordering is $\Delta\chi^2 = 3.6$.
From Fig.~\ref{fig:sq23_sq13_exps} we also notice that, after this combination, the measurement of $\Delta m^2_{31}$ agrees better with the reactor one than in the case of long-baseline data alone. 
However, while the best fit values for normal ordering nearly coincide, there is still a small tension in inverted ordering. Therefore, after the global combination with data from reactor experiments, we obtain the final preference of $\Delta\chi^2 = 6.4$, corresponding to a significance of 2.5$\sigma$.
As for the CP phase $\delta$, the current preference for normal mass ordering is lower than reported in Ref.~\cite{deSalas:2017kay}.
The explanation is the same as before, namely the tension in the combined analysis of T2K and NO$\nu$A for normal mass ordering, due to the different preferred values for $\delta$.
Therefore, any development on this tension will affect the sensitivity of neutrino oscillation data to the mass ordering\footnote{
The weaker preference for normal mass ordering indicated by the 
latest analysis of Super-Kamiokande atmospheric data will affect this result to some extent~\cite{yasuhiro_nakajima_2020_4134680}.}.

\section{Bayesian analysis of neutrino oscillation data}
\label{sec:bayesian}
In this section we turn to the discussion of our Bayesian analysis of neutrino oscillation data.

\subsection{The Bayesian method}
\label{sec:bay_method}
In order to perform a Bayesian analysis of neutrino oscillation data,
we convert the $\chi^2$ functions described in the previous sections
into a likelihood, using the expression
\begin{equation}
\label{eq:llh_chi2}
\ln\mathcal{L}
=
-\frac{\chi^2}{2}\,.
\end{equation}
The analysis is performed using
\texttt{MontePython}~\cite{Audren:2012wb,Brinckmann:2018cvx}
for the computation of the likelihoods and running the Markov Chain Monte Carlo (MCMC) simulations.
\texttt{MontePython} is also used to post-process the MCMC outputs
and obtain the marginalized posteriors and credible intervals.
Where used, Bayesian evidences ($Z$) are computed by means of
\texttt{MCEvidence}~\cite{Heavens:2017afc}.
We checked that \texttt{MCEvidence} provides accurate estimates of the Bayesian evidences
using the nested sampling code \texttt{PolyChord}~\cite{Handley:2015fda,Handley:2015aa},
which is more reliable in case of multivariate distributions but requires longer computation times.
The Bayesian evidences are used in the calculation of the Bayes factors
$B_{\rm NO,IO}=Z_{\rm NO}/Z_{\rm IO}$,
necessary to compare the NO and IO models and decide which of the two is preferred.
The significance in favor of the preferred model is derived
according to Gaussian probabilities as explained in \cite{Gariazzo:2018pei,deSalas:2018bym}.

When considering neutrino mass bounds, discussed in the following section,
the lightest neutrino mass, \mli, is varied in the analyses
using a logarithmic prior in the range $[10^{-3}, 10]$~eV when computing the mass ordering preference
\cite{Gariazzo:2018pei,deSalas:2018bym}
or using a linear prior in the range $[0, 10]$~eV when computing limits
on $\mnu$, $m_\beta$ and $m_{\beta \beta}$
\footnote{
As seen in \cite{Gariazzo:2018pei}, the most efficient prior for sampling the parameter space in order to obtain the mass ordering preference is the logarithmic one,
which gives the same importance to all mass scales. In particular it allows us to sample uniformly the small values of \mli, which are not constrained by neutrino mass probes.
Moreover, linear priors on the neutrino masses can lead to artificially stronger preferences for normal ordering \cite{Gariazzo:2018pei}.
However, using a linear prior on \mli\ is necessary in order to obtain limits on \mnu: i.e.\ 
having a logarithmic prior on \mli\ generates non-trivial distortions to the \mnu\ posterior, which remains much more peaked towards the smaller values of \mnu,
hence precluding a simple comparison with other limits in the literature.
}.

\subsection{Oscillation parameter results}
\label{sec:bay_res}

\begin{figure}[t]
\centering
\includegraphics[width=0.85\textwidth]{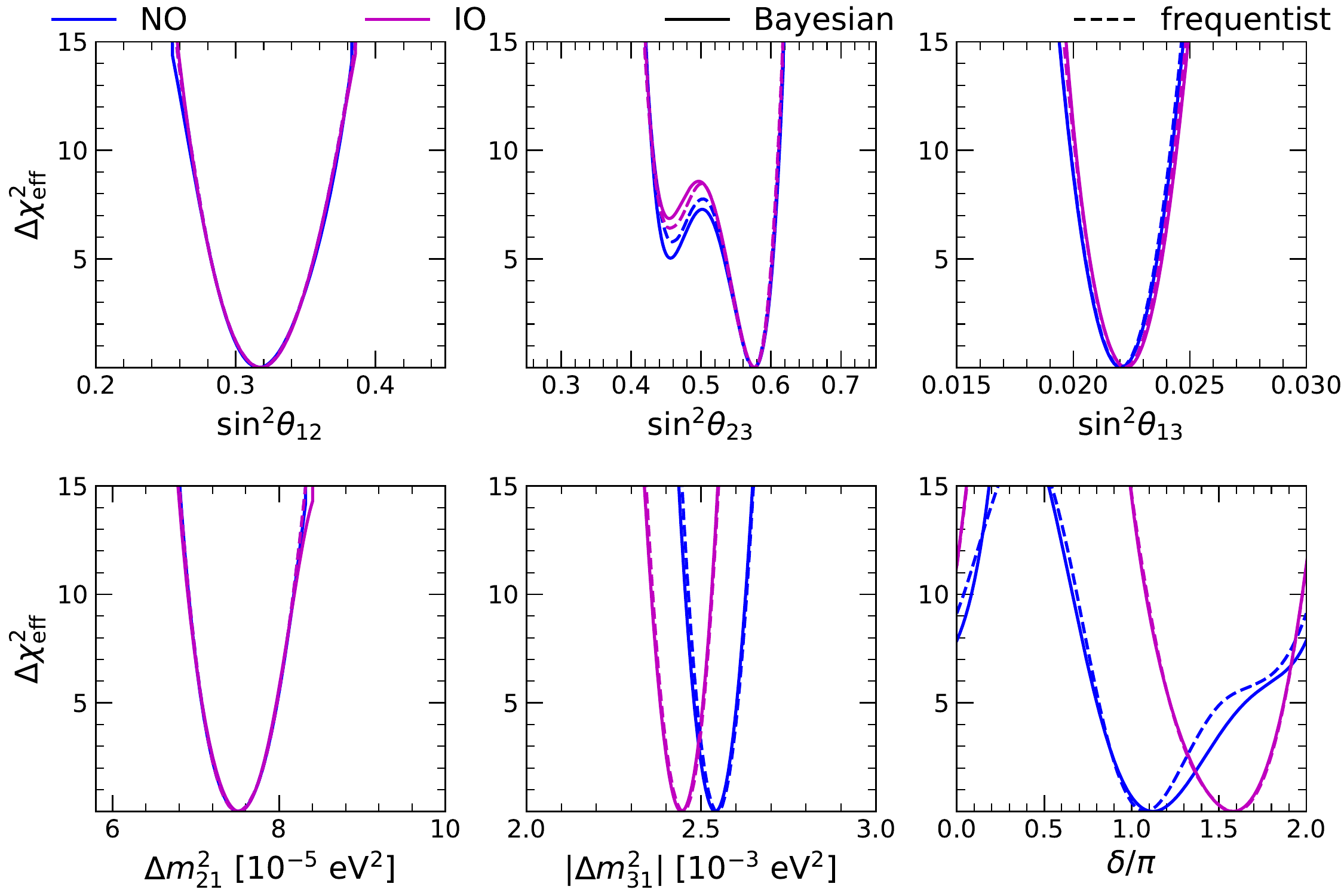}
\caption{
\label{fig:freq_bay}
Summary of neutrino oscillation parameters from our global fit, comparing the Bayesian (solid) and frequentist (dashed)
determinations obtained for normal (blue) and inverted (magenta) ordering.
Note that inverted ordering results are normalized with respect to the minimum $\chi^2$ of inverted ordering.
}
\end{figure}

In order to obtain a Bayesian comparison of the NO and IO spectra, we have to perform numerical analyses which we also use to produce Bayesian neutrino oscillation parameter determinations.
While the likelihood is the same, converted from the $\chi^2$ discussed in the previous sections
according to Eq.~\eqref{eq:llh_chi2}, minor differences appear between the frequentist and Bayesian
analyses, which are shown in Fig.~\ref{fig:freq_bay}.
In the figure, we show the frequentist one-dimensional $\Delta\chi^2$ profiles (dashed lines)
and the marginalized Bayesian posterior probabilities $P(x)$ (solid lines), converted into an effective $\chi^2$ using
\begin{equation}
\label{eq:chi2eff_bay}
\Delta\chi^2_{\rm eff}(x)
=
-2 \log(P(x))\,,
\end{equation}
where $x$ represents any one of the six oscillation parameters.
In the figure, we show NO (blue) and IO (magenta),
normalizing in both cases with respect to the best fit for the same ordering of the spectrum.
Apart for the normalization in the IO case, the dashed lines are the same we show in the global fit summary in Fig.~\ref{fig:chi2_profiles}.
As we can see, most of the posterior distributions are exactly the same as the frequentist profiles.
Minor differences only appear in $\sin^2\theta_{23}$ and $\delta$, but none of the conclusions of the paper are changed.

In the first line of Tab.~\ref{tab:B_NOvsIO} and in Fig.~\ref{fig:bayFactors_NoVsIo},
we report the significance of the Bayesian comparison of NO and IO.
As we can see, the significance decreased slightly with respect to the previous results obtained
in Ref.~\cite{deSalas:2018bym}, due to the mismatch in the determination of $\delta$ by T2K and NO$\nu$A,
as already explained.
Neutrino oscillation data alone give now
$\ln B_{\rm NO,IO} = 3.01 \pm 0.04$,
corresponding to a $2.00
\sigma$ probability for a Gaussian variable.

\section{Absolute scale of neutrino masses}
\label{sec:mass}
Since neutrino oscillations depend only on the mass splittings between the neutrino mass eigenstates,
in order to probe the absolute scale of the neutrino mass, other experiments are required.
In this section, we discuss the status of current probes of the absolute neutrino mass:
kinematic measurements through the observation of the energy spectrum of tritium $\beta$ decay,
neutrinoless double $\beta$ decay, plus cosmological constraints.

\subsection{The end point of $\beta$ decay spectra}  
\label{sec:beta}
The kinematics of $\beta$ decays can be used to probe the absolute scale of neutrino masses.
Depending on the $\beta$-decaying material studied, one can access the mass of neutrinos or antineutrinos,
through measurements of the electron or positron energy spectrum close to the end point.
While, in principle, the electron energy spectrum contains the information encoded in each mass eigenstate, see
e.g.~\cite{Giunti:2007ry}, isolating the individual neutrino masses from such observations is beyond the reach of present experiments.
Current $\beta$ decay probes are only sensitive to the so-called effective electron neutrino mass $m_\beta$,
given by the following sum:
\begin{equation}
\label{eq:m_b}
m_\beta^2
=
\sum_{j=1}^{3} \Uaj{ej} m_{j}^2
\,.
\end{equation}

At the moment, the strongest limits on the effective electron antineutrino mass $m_\beta$
are set by the KATRIN experiment~\cite{Aker:2019uuj}, which obtained the upper limit
$m_\beta<1.1$~eV at 90\% C.L..
This bound applies irrespectively of whether neutrinos are Dirac or Majorana particles.
In our analysis, we do not include data from previous experiments such as MAINZ \cite{Kraus:2004zw} and
TROITSK \cite{Aseev:2012zz}, since they provide much weaker constraints than KATRIN.
When performing the calculations, we take into account the KATRIN results by means of the approximated
analytical likelihood proposed in Eq.~(B.3) of Ref.~\cite{Huang:2019tdh}:
\begin{equation}
\mathcal{L}^{\mathrm{KATRIN}}
\propto
\frac{1}{\sqrt{2\pi}\sigma}
\exp\left(-\frac{1}{2} \left(\frac{m_\beta^2-\mu}{\sigma}\right)^2\right)
\mathrm{erfc}\left(-\frac{\alpha}{\sqrt{2}}\frac{m_\beta^2-\mu}{\sigma}\right)~,
\end{equation}
where erfc is the complementary error function,
$\sigma = 1.506$~eV$^2$,
$\mu= 0.0162$~eV$^2$,
$\alpha=-2.005$
and $m_\beta^2$ is in units of eV$^2$.\\

\subsection{Neutrinoless double $\beta$ decay}
\label{sec:0nbb}

If neutrinos are Majorana particles, one expects that a neutrinoless variety of double beta decay in which no neutrinos are emitted as real particles should take place.
This is called neutrinoless double $\beta$ decay ($0\nu\beta\beta$) and, if it is ever detected, it implies the Majorana nature of neutrinos~\cite{Schechter:1981bd}.
The non-observation of \znbb can then be used to set complementary limits on the neutrino mass scale.
The decay amplitude is given as
\begin{equation}
\label{eq:m_bb}
m_{\beta \beta} = \left| \sum_{j=1}^3 {U_{ej}^2} m_{j} \right|~,
\end{equation}
where $m_j$ are the Majorana masses of the three light neutrinos. Notice the absence of complex conjugation of the lepton mixing matrix elements.
These contain the new CP phases~\cite{Schechter:1980gr,Schechter:1980gk} characteristic of the Majorana neutrinos
(note that the Dirac phase that appears in neutrino oscillations does not appear in the \znbb amplitude).
This is manifest within the original symmetrical parametrization of the lepton mixing matrix~\cite{Schechter:1980gr} in which the Majorana phases
are treated symmetrically, each one associated to the corresponding mixing angles~\footnote{
For a detailed discussion of original ``symmetrical'' phase convention and that of the PDG, see Ref.~\cite{Rodejohann:2011vc}.
The distinction is important if, given a positive \znbb signal, the phases were to be extracted.
For our discussion this subtlety does not matter, as the unknown Majorana phases will be marginalized over.}.

One finds that \znbb probes constrain the half-life $\Thl(\mathcal{N})$ of the isotope involved in the decay
(see e.g.~\cite{DellOro:2016tmg} for a recent review).
Assuming that the dominant mechanism responsible for the $0\nu\beta\beta$ events is light neutrino exchange,
one finds constraints on $\Thl(\mathcal{N})$ which can be translated into bounds on the effective Majorana mass $m_{\beta\beta}$.

The conversion between half-life and effective Majorana mass is 
\begin{equation}
\label{eq:halflife}
\Thl(\mathcal{N})
=
\frac{m_e^2}{\Gps | \Mme |^2 m_{\beta \beta}^2}
\,,
\end{equation}
where $m_e$ is the electron mass,
$\Gps$ is the phase space factor
and $\Mme$ is the nuclear matrix element for the decay.
The latter two terms depend on the isotope under consideration.

Lower limits on the half-life $\Thl(\mathcal{N})$ have been set by various experiments, using different isotopes, including $^{76}$Ge, $^{130}$Te and $^{136}$Xe.
The strongest bounds for these isotopes have been set, respectively, by Gerda \cite{Agostini:2019hzm} for $^{76}$Ge ($\Thl>9\ex{25}$~yr), 
by CUORE \cite{Adams:2019jhp} for $^{130}$Te ($\Thl>3.2\ex{25}$~yr) and by KamLAND-Zen \cite{KamLAND-Zen:2016pfg} for $^{136}$Xe ($\Thl>1.07\ex{26}$~yr), all at 90\% C.L.
In our analyses, we consider bounds from the above-mentioned experiments using approximated analytical expressions for the three likelihoods:
\begin{eqnarray}
-\ln\mathcal{L}^{\mathrm{Gerda}}
&\propto&
-5.5 + 26.7 \,(\Thl)^{-1} + 38.4\,(\Thl)^{-2}
\,,\\
-\ln\mathcal{L}^{\mathrm{CUORE}}
&\propto&
4.02 + 10.5 \,(\Thl)^{-1} + 8.6 \,(\Thl)^{-2}
\,,\\
-\ln\mathcal{L}^{\mathrm{KamLAND-Zen}}
&\propto&
9.71 \,(\Thl)^{-1} + 28.1 \,(\Thl)^{-2}
\,.
\end{eqnarray}
The latter expression was proposed in \cite{Caldwell:2017mqu}, while the former two have been obtained using the information
from \cite{Agostini:2019hzm} and \cite{Adams:2019jhp}, respectively, and using the general fitting formula proposed in \cite{Caldwell:2017mqu}.

When performing analyses including constraints from neutrinoless double $\beta$ decay probes,
we marginalize over the two Majorana phases in their allowed range.
Moreover, in order to take into account the theoretical uncertainties in the calculation
of the nuclear matrix elements, we vary them within the ranges
\begin{eqnarray}
\mathcal{M}_{0 \nu}^{^{76}\mathrm{Ge}}  &\in& [3.35,5.75]\,,\\
\mathcal{M}_{0 \nu}^{^{130}\mathrm{Te}} &\in& [1.75,5.09]\,,\\
\mathcal{M}_{0 \nu}^{^{136}\mathrm{Xe}} &\in& [1.49,3.69]\,,
\end{eqnarray}
which correspond to the proposed $1\sigma$ range from~\cite{Vergados:2016hso} (see Tab.~6 in that reference).

This way, we find that the bounds on the half-life $\Thl(\mathcal{N})$ from the three experiments under consideration imply the following upper limits on the effective mass:
$m_{\beta\beta}<104-228$~meV by Gerda \cite{Agostini:2019hzm},
$m_{\beta\beta}<75-350$~meV by CUORE \cite{Adams:2019jhp}
and $m_{\beta\beta}<61-165$~meV by KamLAND-Zen \cite{KamLAND-Zen:2016pfg}, respectively,
where the lower (upper) values correspond to the most aggressive (conservative) choices for the nuclear matrix elements.

\subsection{Cosmological probes}
\label{sec:cosmo}

Stronger, though model-dependent (see e.g.~\cite{Gariazzo:2018meg}), are the limits on the sum of the neutrino masses provided by cosmological observations.
They arise mainly from the combination of Cosmic Microwave Background (CMB) and Baryon Acoustic Oscillation (BAO) measurements (see e.g.~\cite{Lattanzi:2017ubx}).
Due to the anticorrelation between the sum of the neutrino masses and the Hubble parameter, it is also interesting to consider constraints on the latter quantity.

In our analyses, we consider the most recent observations of the CMB spectrum by
Planck \cite{Akrami:2018vks,Aghanim:2018eyx},
which measures the temperature and polarization spectra in a wide range of multipoles
through the respective two-point correlation functions \cite{Aghanim:2019ame}
and the lensing \cite{Aghanim:2018oex} potential through the
four-point correlation function.
We include BAO observations from the
6dF \cite{Beutler:2011hx},
SDSS DR7 Main Galaxy Sample (MGS) \cite{Ross:2014qpa}
and BOSS DR12 \cite{Alam:2016hwk} galaxy redshift surveys.
Bounds on the expansion of the universe quantified by the Hubble parameter $H(z)$ also come from measurements at $z=0.45$ \cite{Moresco:2016mzx}.
Constraints from observations of Type Ia Supernovae are also taken into account, by means of the Pantheon sample \cite{Scolnic:2017caz}.
Here we will indicate by ``Cosmo'' the combination that includes Planck CMB temperature,
polarization and lensing spectra, BAO measurements,
$H(z)$ observations and Supernovae luminosity distance data.
Finally, the most recent determination of the Hubble parameter today,
$H_0 = 74.03 \pm 1.42$~km/s/Mpc from \cite{Riess:2019cxk}, is also
included in some cases in order to illustrate the impact of the $\mnu- H_0$ degeneracy.

The calculation of predicted cosmological observables is performed
using the Boltzmann solver code
\texttt{CLASS}~\cite{Lesgourgues:2013bra,Blas:2011rf,Lesgourgues:2011re}.
Our fiducial cosmological model is a minimal extension of the $\Lambda$CDM model,
which is described by the usual six free parameters. Namely, the
baryon and cold dark matter physical densities $\Omega_bh^2$ and $\Omega_ch^2$,
the angular size of the sound horizon at last-scattering $\theta_s$,
the optical depth to reionization $\tau$ and the amplitude and tilt of
the primordial scalar power spectrum $A_s$ and $n_s$. We fix the number of ultra-relativistic species to zero and we add three massive neutrinos,
each with its own mass. Such masses are derived from the lightest
neutrino mass \mli\ and the two mass splittings
before performing the cosmological calculations. The total neutrino mass
in the two orderings reads as
\begin{equation}
\begin{split}
\mnu^{\rm{NO}} &= \mli + \sqrt{m_{1}^{2}+\Delta m^{2}_{21}}+\sqrt{m_{1}^{2}+\Delta m^{2}_{31}}~, \\
\mnu^{\rm{IO}} &= \mli + \sqrt{m_{3}^{2}+|\Delta m^{2}_{31}|}+\sqrt{m_{3}^{2}+|\Delta m^{2}_{31}|+\Delta m^{2}_{21}}~.
\end{split}
\end{equation}

\subsection{Global results on the neutrino mass scale and mass ordering}
\label{sec:globmass}

\begin{figure}[t]
\centering
\includegraphics[width=0.7\textwidth]{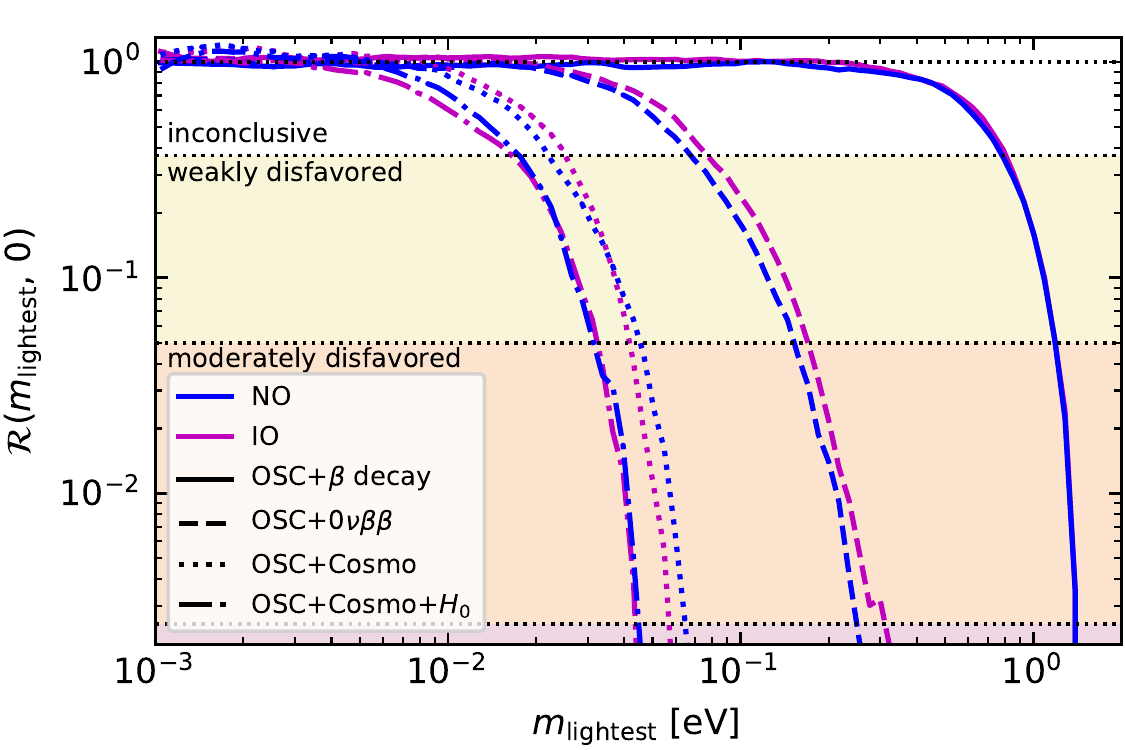}
\caption{
\label{fig:R_func}
Prior-independent \cite{Gariazzo:2019xhx} constraints
on the lightest neutrino mass from different data combinations,
for normal (blue) and inverted (magenta) neutrino mass ordering.
}
\end{figure}

Once we add in our analyses the constraints from $\beta$ decay, neutrinoless double-$\beta$ decay
and cosmology, we are able to obtain upper bounds on the absolute scale of neutrino masses.
Here we present the results in terms of \mli,
which is the quantity we can compare in an easier way when discussing the various probes.
In Fig.~\ref{fig:R_func} we report the constraints on \mli\ in a prior-independent way,
using the method of
Refs.~\cite{Astone:1999wp,DAgostini:2000edp,DAgostini:2003} and recently revived in \cite{Gariazzo:2019xhx}.
The plotted function
\begin{equation}
\label{eq:R_func}
\mathcal{R}(\mli,0)
\equiv
\frac{p(\mli)/\pi(\mli)}{p(\mli=0)/\pi(\mli=0)}
\end{equation}
shows the ratio between the posterior $p(\mli)$ and the prior $\pi(\mli)$
normalized with respect to the same ratio computed at $\mli=0$,
for different data sets, and comparing normal (blue) to inverted (magenta) mass ordering.
Such quantity, which has the property of being independent of the shape and normalization
of the prior $\pi(\mli)$, is statistically equivalent to a Bayes factor between
a model where  $\mli$ has been fixed to some value
and one where \mli\ is equal to zero.
Since the considered measurements are insensitive to the value of \mli\ when it is very small,
the function $\mathcal{R}$ is expected to be equal to one for small \mli~%
\footnote{
Note that, due to the numerical noise in the MCMC,
the posterior at small values of \mli\ is not perfectly stable
and, as a consequence, $\mathcal{R}$ is not exactly constant.
},
while it decreases when large values of \mli\ become disfavored.
In the same way as a Bayes factor, we can compare the constraining power of different data sets by means of the Jeffreys' scale.
The horizontal lines in Fig.~\ref{fig:R_func} show the values $\ln\mathcal{R}=-1, -3, -6$, which separate regions where the significance is none, weak, moderate and strong,
according to the Jeffreys' scale we adopt.

Analyzing the figure, it is possible to notice that the results obtained from the $\beta$-decay data (solid)
are completely insensitive to the mass ordering, and provide the weakest constraints on \mli\ nowadays.
We must remember, however, that $\beta$ decay measurements provide the
most robust, less biased constraints
on the absolute scale of neutrino masses, as they are completely model independent.
Neutrinoless double-$\beta$ decay bounds (dashed), which only apply to Majorana neutrinos,
provide stronger bounds, with minor differences between NO and IO, which are just due to MCMC noise.
The stronger constraints on \mli\ come from cosmological measurements.
The inclusion of a prior on $H_0$~\cite{Riess:2019cxk} (dashed-dotted)
further reduces the allowed range for \mli\
with respect to the Cosmo data set (dotted). This is a consequence of the anticorrelation between \mnu\ and $H_0$,
which reduces the allowed range for \mnu\
when $H_0$ is forced to increase by
the tension between the local measurement and the value of $H_0$ derived
from CMB observations, see e.g.~\cite{Lattanzi:2017ubx}.
To summarize, the OSC+Cosmo fit strongly disfavors values of \mli\ above
0.065\unskip~eV
(0.058\unskip~eV),
while in the OSC+Cosmo+$H_0$ fit the strongly disfavored values above
0.045\unskip~eV
for both mass orderings.
For an easier comparison with bounds existing in literature,
we also quote the marginalized limits on the sum of the neutrino masses at $2\sigma$,
computed with a linear prior on \mli,
which are
0.15\unskip~eV
(0.16\unskip~eV)
when considering OSC+Cosmo data,
while in the OSC+Cosmo+$H_0$ fit they become
0.12\unskip~eV
(0.15\unskip~eV),
for NO (IO).
Notice that the NO and IO bounds differ,
and in both cases these results are less constraining than
those obtained by the Planck collaboration and in 
Refs.~\cite{Vagnozzi:2018jhn,Vagnozzi:2017ovm,Giusarma:2016phn}
after considering very similar cosmological observations.
Namely, in Ref.~\cite{Aghanim:2018eyx} it
is quoted $\mnu <0.12$~eV from CMB temperature, polarization,
lensing and BAO observations.
This is due to the different lower prior
assumed in our global fit: while the Planck collaboration just assumes
a physical prior on the sum of the neutrino masses,
i.e.\ $\mnu>0$,
here the lower value of the prior is determined by neutrino
oscillation experiments assuming \mli$=0$, and it is therefore
different for NO and IO. We refer to reader to
Ref.~\cite{Vagnozzi:2017ovm} and Ref.~\cite{Gariazzo:2018meg} for an assessment of the
changes in the upper bounds on the total neutrino mass after taking
into account neutrino oscillation information and
the uncertainty on the underlying cosmological model, respectively.

\begin{figure}[t]
\centering
\includegraphics[width=0.5\textwidth]{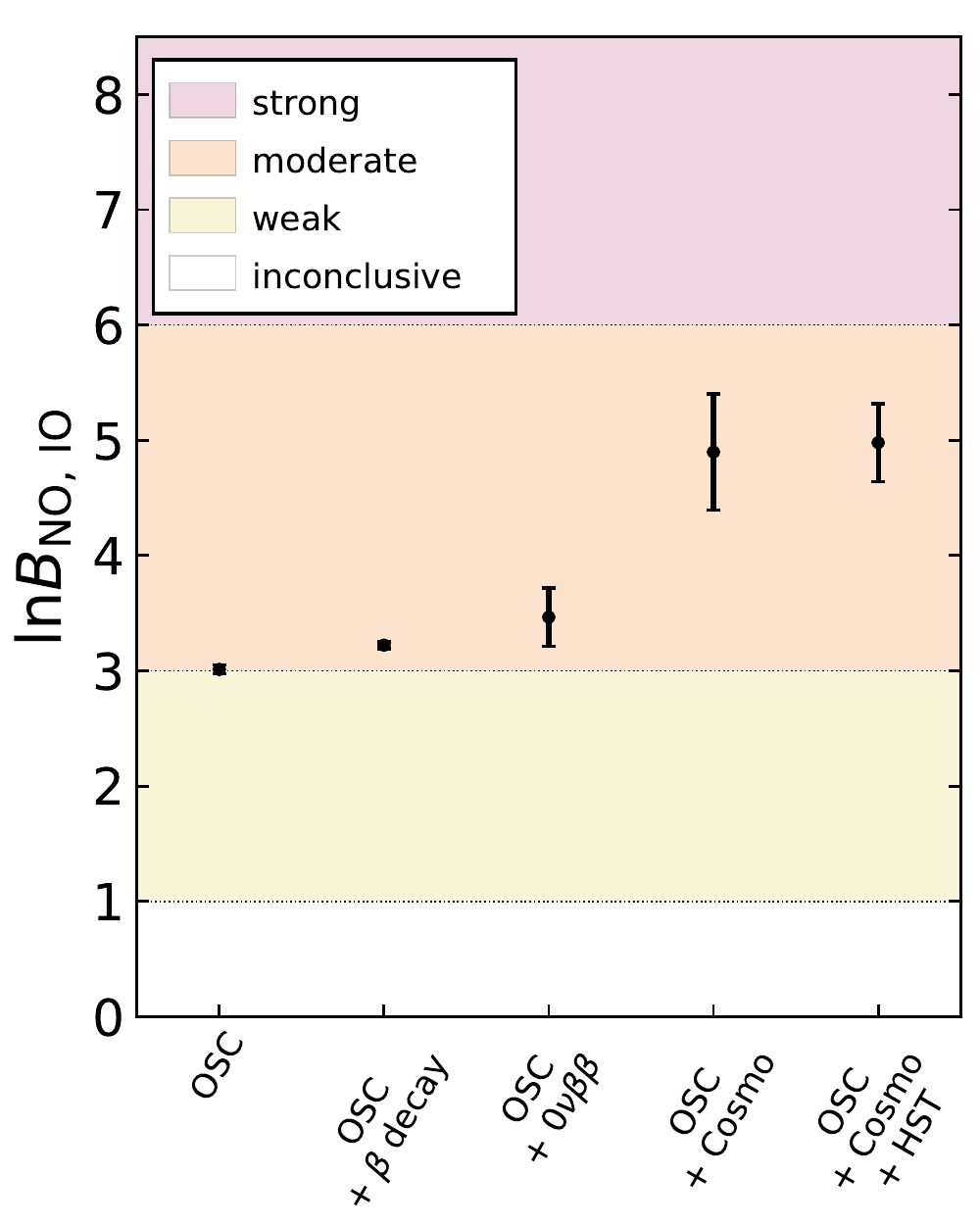}
\caption{
\label{fig:bayFactors_NoVsIo}
Bayes factors comparing normal and inverted neutrino mass ordering,
using oscillation data alone and in combination with other data sets
sensitive to the absolute scale of neutrino masses.
}
\end{figure}

\begin{table}[t]
\centering
\begin{tabular}{|c|c|c|}
\hline
data set & ~~~~$\ln B_{\rm NO,IO}$~~~~ & ~~~$N\sigma$~~~\\
\hline
OSC                    & $3.01 \pm 0.04$    & 2.00
 \\
OSC + $\beta$ decay    & $3.22 \pm 0.03$  & 2.07 \\
OSC + $0\nu\beta\beta$ & $3.46 \pm 0.25$ & 2.17 \\
OSC + Cosmo            & $4.90 \pm 0.50$  & 2.68 \\
OSC + Cosmo + $H_0$    & $4.98 \pm 0.34$ & 2.70 \\
\hline
\end{tabular}
\caption{
\label{tab:B_NOvsIO}
Bayes factors and significance in terms of standard errors
of normal versus inverted mass ordering for various data combinations.
}
\end{table}

Considering absolute neutrino mass measurements also affects the
preference in favor of NO previously reported from oscillation data only.
While the strength of $\beta$ decay constraints is not yet sufficient to discriminate the mass ordering,
in the case of Majorana neutrinos, the bounds obtained by $0\nu\beta\beta$ experiments
provide some additional significance in favor of NO,
from $2.00
\sigma$ (oscillations only)
to $2.17\sigma$, as reported in Tab.~\ref{tab:B_NOvsIO}
and Fig.~\ref{fig:bayFactors_NoVsIo}.
The significance increases even more when the constraints on the neutrino mass from cosmology
are taken into account.
In such case, we obtain a preference of $\sim 2.68\sigma$
when considering oscillation data plus the Cosmo set,
which does not vary significantly if a prior on the Hubble parameter is also included.
The stronger preference obtained in favor of NO in such cases is due to the fact
that cosmology puts stronger constraints on the absolute scale of neutrino masses.
In the IO case, stronger bounds on \mnu\ result in a smaller available parameter space volume,
since the minimum allowed value for \mnu\ is bounded from below by $\sim0.1$~eV
instead of the $\sim0.06$~eV that apply for NO,
which is therefore preferred.

\begin{figure}
\includegraphics[width=0.9\textwidth]{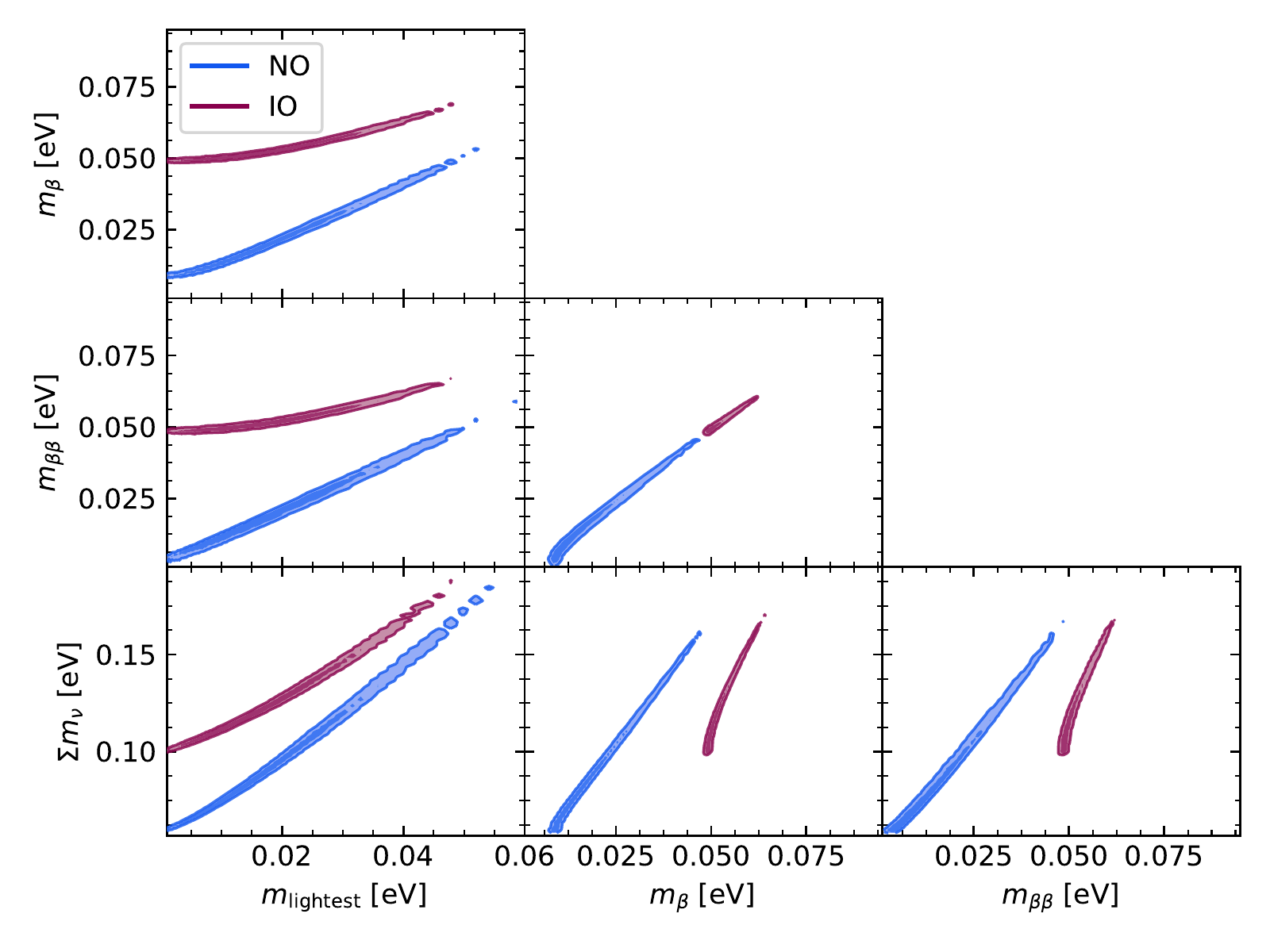}
\caption{
\label{fig:masses_cosmo}
Marginalized allowed regions at $1\sigma$ (dark area) and $2\sigma$ (light area)
for \mli, $m_\beta$, $m_{\beta \beta}$ and \mnu,
obtained considering the data combination we denote as OSC+Cosmo,
for NO (blue) and IO (magenta).}
\end{figure}

Finally, we report in Fig.~\ref{fig:masses_cosmo} the allowed regions at $1,\,2\sigma$
for the mass parameters \mli, $m_\beta$, $m_{\beta \beta}$ and \mnu\ 
obtained considering the OSC+Cosmo data set.
We do not show the regions allowed by $\beta$-decay and
$0\nu\beta\beta$ experiments as they are outside the scale of the respective effective parameter:
in other words, the constraints that we obtain considering cosmological data are much tighter
than those obtained from terrestrial experiments (see Secs.~\ref{sec:beta} and \ref{sec:0nbb}).

\section{Summary of the global fit}
\label{sec:conc}

\begin{figure}[p]
 \centering
\includegraphics[width=0.65\textwidth]{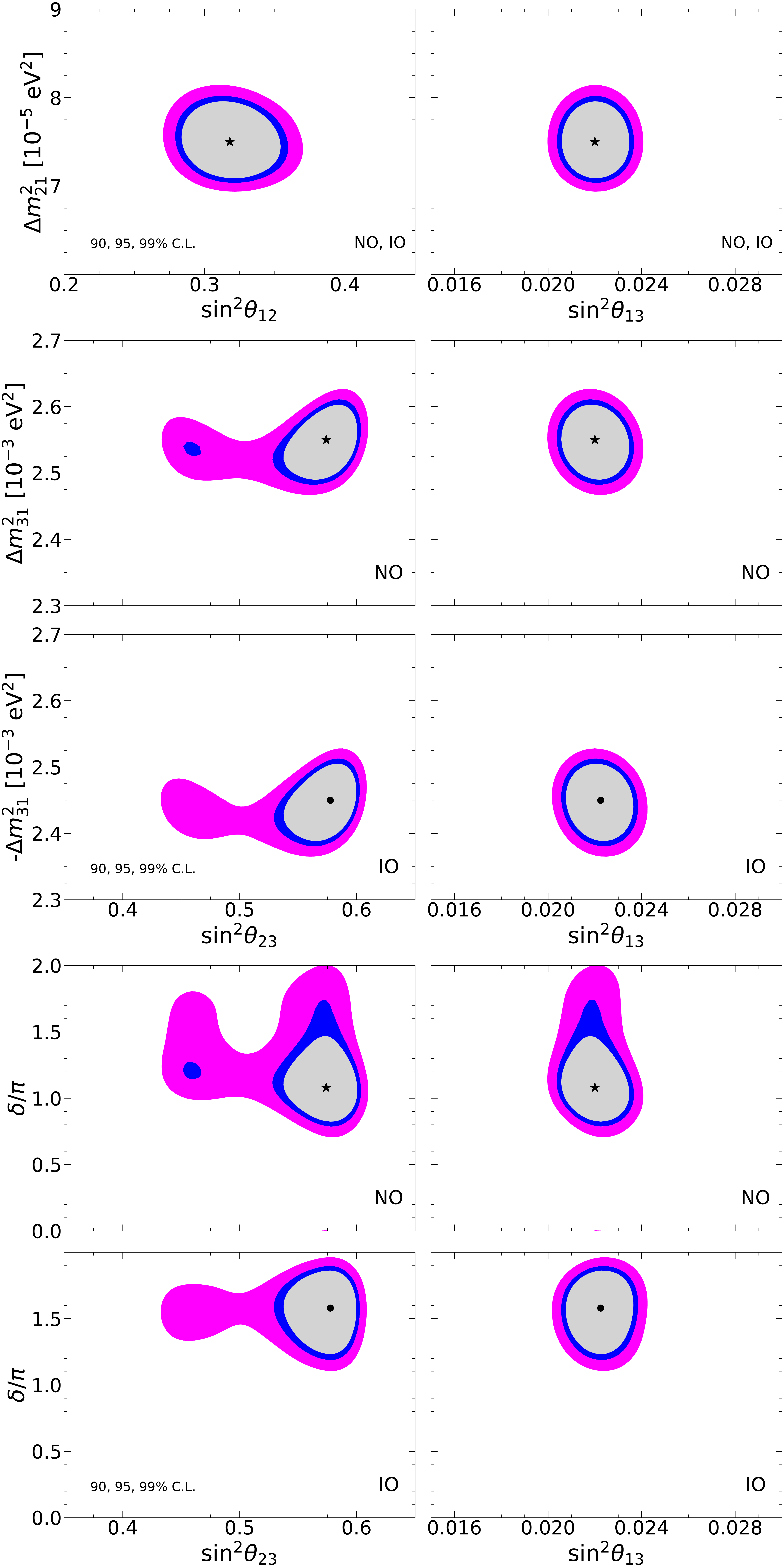}
\caption{Overall summary of our global fit to neutrino oscillation data. Global fit regions correspond to 90, 95 and 99\% C.L.\ (2 d.o.f.).
Regions for inverted ordering are plotted with respect to the local minimum in this neutrino mass ordering. The absolute minimum corresponding to NO is indicated by a star, while the local minimum in IO is denoted by a black dot.}
\label{fig:glob_2dim}
\end{figure}
\begin{figure}[t!]
\centering
\includegraphics[width=1\textwidth]{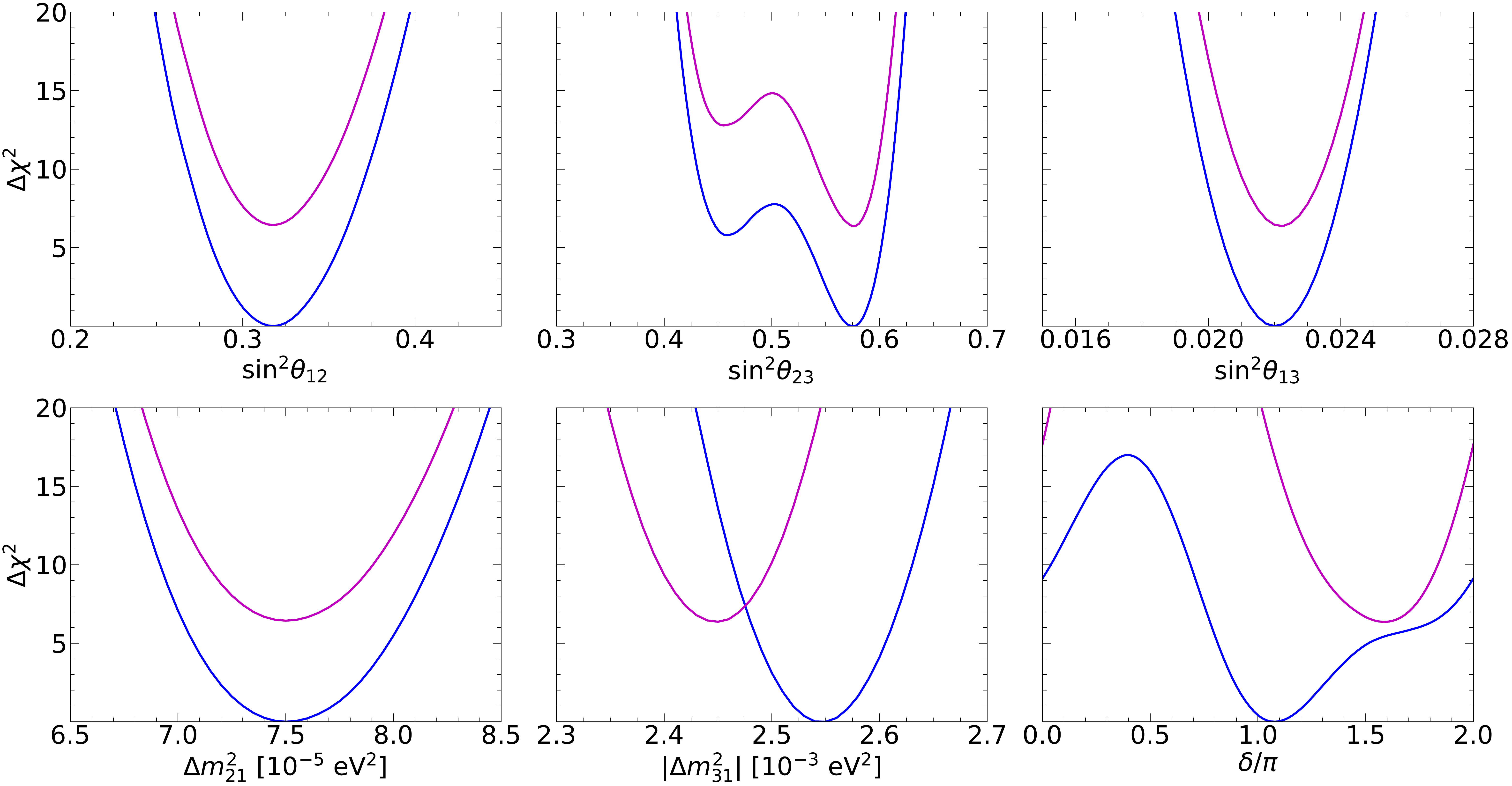}
\caption{
\label{fig:chi2_profiles}
 Overall summary of neutrino oscillation parameter determinations. Blue lines correspond to NO and magenta lines to IO.}
\end{figure}
\begin{table}[t!]\centering
\catcode`?=\active \def?{\hphantom{0}}
\begin{tabular}{|c|ccc|}
\hline
parameter & best fit $\pm$ $1\sigma$ & \hphantom{x} 2$\sigma$ range \hphantom{x} & \hphantom{x} 3$\sigma$ range \hphantom{x}
\\ \hline
$\Delta m^2_{21} [10^{-5}$eV$^2$]  &  $7.50^{+0.22}_{-0.20}$  &  7.12--7.93  &  6.94--8.14  \\[3mm]
$|\Delta m^2_{31}| [10^{-3}$eV$^2$] (NO)  &  $2.55^{+0.02}_{-0.03}$  &  2.49--2.60  &  2.47--2.63  \\
$|\Delta m^2_{31}| [10^{-3}$eV$^2$] (IO)  &  $2.45^{+0.02}_{-0.03}$  &  2.39--2.50  &  2.37--2.53  \\[3mm]
$\sin^2\theta_{12} / 10^{-1}$         &  $3.18\pm0.16$  &  2.86--3.52  &  2.71--3.69  \\
$\theta_{12} /\degree$                &  $34.3\pm1.0$  &  32.3--36.4  &  31.4--37.4
\\[3mm]
$\sin^2\theta_{23} / 10^{-1}$       (NO)  &  $5.74\pm0.14$  &  5.41--5.99  &  4.34--6.10  \\
$\theta_{23} /\degree$              (NO)  &  $49.26\pm0.79$  &  47.37--50.71  &  41.20--51.33  \\
$\sin^2\theta_{23} / 10^{-1}$       (IO)  &  $5.78^{+0.10}_{-0.17}$  &  5.41--5.98  &  4.33--6.08  \\
$\theta_{23} /\degree$              (IO)  &  $49.46^{+0.60}_{-0.97}$  &  47.35--50.67  &  41.16--51.25  \\[3mm]
$\sin^2\theta_{13} / 10^{-2}$       (NO)  &  $2.200^{+0.069}_{-0.062}$  &  2.069--2.337  &  2.000--2.405  \\
$\theta_{13} /\degree$              (NO)  &  $8.53^{+0.13}_{-0.12}$  &  8.27--8.79  &  8.13--8.92  \\
$\sin^2\theta_{13} / 10^{-2}$       (IO)  &  $2.225^{+0.064}_{-0.070}$  &  2.086--2.356  &  2.018--2.424  \\
$\theta_{13} /\degree$              (IO)  &  $8.58^{+0.12}_{-0.14}$  &  8.30--8.83  &  8.17--8.96  \\[3mm]
$\delta/\pi$                        (NO)  &  $1.08^{+0.13}_{-0.12}$  &  0.84--1.42  &  0.71--1.99  \\
$\delta/\degree$                    (NO)  &  $194^{+24}_{-22}$  &  152--255  &  128--359  \\
$\delta/\pi$                        (IO)  &  $1.58^{+0.15}_{-0.16}$  &  1.26--1.85  &  1.11--1.96  \\
$\delta/\degree$                    (IO)  &  $284^{+26}_{-28}$  &  226--332  &  200--353  \\[3mm]

\hline
\end{tabular}
\caption{
Neutrino oscillation parameters summary determined from the global analysis.
The ranges for inverted ordering refer to the local minimum for this neutrino mass ordering.}
\label{tab:sum-2020}
\end{table}

In this study we have first analyzed global data coming only from neutrino oscillation experiments. These hold in complete generality.
In a second step, we have combined the oscillation results with direct neutrino mass probes such as $\beta$ decay, neutrinoless double $\beta$ decay and cosmological observations.
The latter has its caveats, for example, \znbb restrictions apply only to Majorana neutrinos, while cosmological considerations suffer from a higher degree of model-dependence.

The results from our frequentist global fit to neutrino oscillation data are summarized in Figs.~\ref{fig:glob_2dim} and~\ref{fig:chi2_profiles} and in Tab.~\ref{tab:sum-2020}.
We have reanalyzed SNO data obtaining now a more constraining upper bound on the solar mixing angle $\sin^2\theta_{12}$ than that obtained in the recent global fit, Ref.~\cite{deSalas:2017kay},
with the new best fit value slightly smaller.
While the determination of the solar mass splitting $\Delta m_{21}^2$ remains unchanged, the best fit value is also smaller than that obtained before.
Due to new short-baseline reactor data from Daya Bay and RENO, we obtain an improved measurement of $\sin^2\theta_{13}$ and a larger best fit value of $\sin^2\theta_{13}=0.0220$.
Also the best fit value of the atmospheric mass splitting $\Delta m_{31}^2$ is now larger, with a slightly better determination.
Regarding the atmospheric angle, we obtain the best fit value $\sin^2\theta_{23} = 0.574$ (0.578) for normal (inverted) ordering.
Indeed, the preference for the second octant obtained here is stronger than in Ref.~\cite{deSalas:2017kay}, and lower octant solutions are now disfavored with
$\Delta\chi^2 \geq 5.8$ (6.4) for normal (inverted) ordering.

However, for the case of the CP-violating phase $\delta$, we obtain a weaker result in comparison with our previous global fit~\cite{deSalas:2017kay}, due to the
mismatch in the value of $\delta$ extracted by T2K and NO$\nu$A. 
The best fit is obtained for $\delta = 1.08\pi$ (1.58$\pi$) for normal (inverted) ordering. 
Concerning the CP-conserving values, $\delta = 0$ is disfavored with $\Delta\chi^2 = 9.1$ (11.3) for NO (IO), while $\delta = \pi$,
remains allowed with $\Delta\chi^2 = 0.4$ for NO and it is excluded with $\Delta\chi^2 =14.6$ in IO. 
This is due to the fact that the aforementioned mismatch in the extracted value of $\delta$ by the T2K and NO$\nu$A experiments only occurs
for normal ordering. 
Indeed, for inverted mass ordering both experiments prefer values close to maximal CP violation, with $\delta \approx 1.5\pi$. 
Finally, the very same mismatch reduces the statistical significance of the preference for NO from the 3.4$\sigma$ obtained in
Ref.~\cite{deSalas:2017kay} to the 2.5$\sigma$ derived in the frequentist analysis presented in this work.

Our results from the Bayesian analysis are summarized in Fig.~\ref{fig:bayFactors_NoVsIo} and Tab.~\ref{tab:B_NOvsIO}.
Performing the Bayesian analysis, we obtain a Bayes factor of $\ln B =
3.01 \pm 0.04$ in favor of normal neutrino
mass ordering from neutrino oscillation data alone. This would
correspond to a Gaussian preference of
$2.00
\sigma$. The determination of
neutrino oscillation parameters shows also an excellent agreement among the frequentist and Bayesian analyses.

While the inclusion of $\beta$-decay data basically does not change the former Bayes factor, data from $0\nu\beta\beta$ experiments mildly increase that figure to
$\ln B = 3.46 \pm 0.25$, now indicating moderate preference for normal neutrino mass ordering. 
Note, however, that this improvement would only apply if neutrinos are Majorana particles. 
The combination with data from cosmological observations is independent of the neutrino nature and leads to a Bayes factor of $\ln B = 4.90 \pm 0.50$, still moderately preferring normal neutrino mass ordering and corresponding to a statistical significance of $2.68\sigma$.
Finally, when we also include in the cosmological observations a prior
on the Hubble constant we obtain $\ln B=4.98 \pm 0.34$, corresponding to a
preference for NO of $2.70\sigma$.

Concerning the cosmological limits on the sum of the neutrino masses, the tightest $2\sigma$ bound we obtain here is $\sum
m_\nu < 0.12$ ($0.15$)~eV for NO (IO) taking into account CMB temperature, polarization and lensing measurements
from the Planck satellite, BAO observations, $H(z)$ information and
Supernovae Ia data. These limits are slightly weaker than those existing
in the literature due to our prior, that takes into account neutrino
oscillation results as an input. 

Overall, we have seen that the determination of some of the neutrino parameters has improved thanks to new oscillation data, while the
determination of $\delta$ and the neutrino mass ordering has worsened, due to a new tension in current long-baseline accelerator measurements.
We have also seen that the inclusion of non-oscillation data, especially from cosmological observations, enhances the preference for
normal neutrino mass ordering from ``almost weak'' to ``moderate''.

In summary, neutrino oscillation parameters are currently measured with very good precision.
In the upcoming years these accurate measurements will further improve, allowing for better sensitivities to New Physics effects,
which may show up as sub-leading effects in the neutrino oscillation probabilities.

\section*{Acknowledgments}
We thank Chad Finley from the IceCube collaboration for useful comments on the IceCube DeepCore analysis.
We are very grateful to Soo-Bong Kim from the RENO collaboration for providing the digitized data for 2900 days of running time.
Work supported by the Spanish grants FPA2017-85216-P and FPA2017-85985-P (AEI/FEDER, UE), PROMETEO/2018/165 and PROMETEO/2019/083 (Generalitat Valenciana) and the Red Consolider MultiDark FPA2017-90566-REDC.
PFdS acknowledges support by the Vetenskapsr{\aa}det (Swedish Research Council) through contract No.\ 638-2013-8993 and the Oskar Klein Centre for Cosmoparticle Physics.
DVF acknowledges the financial support of UdeM under the internal call, number 48, 
``proyectos de investigación en colaboración" project 1117 and thanks A.\ Tapia for useful 
discussions.
SG acknowledges financial support by the ``Juan de la Cierva-Incorporaci\'on'' program (IJC2018-036458-I) of the Spanish MINECO, until September 2020, and by the European Union's Horizon 2020 research and innovation programme under the Marie Skłodowska-Curie grant agreement No 754496 (project FELLINI) starting from October 2020.
PMM acknowledges financial support from the FPU grant FPU18/04571.
OM is also supported by the European Union Horizon 2020 research and innovation program (grant agreements No. 690575 and 67489).
CAT is supported by the FPI grant BES-2015-073593 and by the research grant
``The Dark Universe: A Synergic Multimessenger Approach'' number 2017X7X85K
under the program ``PRIN 2017'' funded by the Ministero dell'Istruzione,
Universit\`a e della Ricerca (MIUR).
MT acknowledges financial support from MINECO through the Ram\'{o}n y
Cajal contract RYC-2013-12438.


\providecommand{\href}[2]{#2}\begingroup\raggedright\endgroup


\begin{thebibliography}{100}

\bibitem{deSalas:2017kay}
P.~F. de~Salas, D.~V. Forero, C.~A. Ternes, M.~T{\'o}rtola, and J.~W.~F. Valle,
  ``{Status of neutrino oscillations 2018: 3 $\sigma$ hint for normal mass
  ordering and improved {CP} sensitivity},''
  \href{http://dx.doi.org/10.1016/j.physletb.2018.06.019}{{\em Phys.\ Lett.\ B}
  {\bfseries 782} (Jul, 2018) 633--640},
  \href{http://arxiv.org/abs/1708.01186}{{\ttfamily arXiv:1708.01186
  [hep-ph]}}.

\bibitem{Forero:2014bxa}
D.~V. Forero, M.~Tortola, and J.~W.~F. Valle, ``{Neutrino oscillations
  refitted},'' \href{http://dx.doi.org/10.1103/PhysRevD.90.093006}{{\em
  Phys.Rev.D} {\bfseries 90} no.~9, (2014) 093006},
  \href{http://arxiv.org/abs/1405.7540}{{\ttfamily arXiv:1405.7540 [hep-ph]}}.

\bibitem{Tortola:2012te}
D.~V. Forero, M.~Tortola, and J.~W.~F. Valle, ``{Global status of neutrino
  oscillation parameters after Neutrino-2012},''
  \href{http://dx.doi.org/10.1103/PhysRevD.86.073012}{{\em Phys. Rev.}
  {\bfseries D86} (2012) 073012},
  \href{http://arxiv.org/abs/1205.4018}{{\ttfamily arXiv:1205.4018 [hep-ph]}}.

\bibitem{Schwetz:2011zk}
T.~Schwetz, M.~Tortola, and J.~W.~F. Valle, ``{Where we are on $\theta_{13}$:
  addendum to `Global neutrino data and recent reactor fluxes: status of
  three-flavour oscillation parameters'},''
  \href{http://dx.doi.org/10.1088/1367-2630/13/10/109401}{{\em New J. Phys.}
  {\bfseries 13} (2011) 109401},
  \href{http://arxiv.org/abs/1108.1376}{{\ttfamily arXiv:1108.1376 [hep-ph]}}.

\bibitem{Schwetz:2011qt}
T.~Schwetz, M.~Tortola, and J.~W.~F. Valle, ``{Global neutrino data and recent
  reactor fluxes: status of three-flavour oscillation parameters},''
  \href{http://dx.doi.org/10.1088/1367-2630/13/6/063004}{{\em New J. Phys.}
  {\bfseries 13} (2011) 063004},
  \href{http://arxiv.org/abs/1103.0734}{{\ttfamily arXiv:1103.0734 [hep-ph]}}.

\bibitem{Schwetz:2008er}
T.~Schwetz, M.~Tortola, and J.~W.~F. Valle, ``{Three-flavour neutrino
  oscillation update},''
  \href{http://dx.doi.org/10.1088/1367-2630/10/11/113011}{{\em New J. Phys.}
  {\bfseries 10} (2008) 113011},
  \href{http://arxiv.org/abs/0808.2016}{{\ttfamily arXiv:0808.2016 [hep-ph]}}.

\bibitem{Maltoni:2004ei}
M.~Maltoni, T.~Schwetz, M.~A. Tortola, and J.~W.~F. Valle, ``{Status of global
  fits to neutrino oscillations},''
  \href{http://dx.doi.org/10.1088/1367-2630/6/1/122}{{\em New J.Phys.}
  {\bfseries 6} (2004) 122},
  \href{http://arxiv.org/abs/hep-ph/0405172}{{\ttfamily arXiv:hep-ph/0405172
  [hep-ph]}}.

\bibitem{Maltoni:2003da}
M.~Maltoni, T.~Schwetz, M.~A. Tortola, and J.~W.~F. Valle, ``{Status of three
  neutrino oscillations after the SNO salt data},''
  \href{http://dx.doi.org/10.1103/PhysRevD.68.113010}{{\em Phys. Rev.}
  {\bfseries D68} (2003) 113010},
  \href{http://arxiv.org/abs/hep-ph/0309130}{{\ttfamily arXiv:hep-ph/0309130
  [hep-ph]}}.

\bibitem{Esteban:2020cvm}
I.~Esteban, M.~Gonzalez-Garcia, M.~Maltoni, T.~Schwetz, and A.~Zhou, ``{The
  fate of hints: updated global analysis of three-flavor neutrino
  oscillations},'' \href{http://dx.doi.org/10.1007/JHEP09(2020)178}{{\em JHEP}
  {\bfseries 09} (2020) 178}, \href{http://arxiv.org/abs/2007.14792}{{\ttfamily
  arXiv:2007.14792 [hep-ph]}}.

\bibitem{Capozzi:2020qhw}
F.~Capozzi, E.~Di~Valentino, E.~Lisi, A.~Marrone, A.~Melchiorri, and
  A.~Palazzo, ``{Addendum to: Global constraints on absolute neutrino masses
  and their ordering},'' \href{http://arxiv.org/abs/2003.08511}{{\ttfamily
  arXiv:2003.08511 [hep-ph]}}.

\bibitem{Ahmad:2002jz}
{\bfseries SNO} Collaboration, Q.~Ahmad {\em et~al.}, ``{Direct evidence for
  neutrino flavor transformation from neutral current interactions in the
  Sudbury Neutrino Observatory},''
  \href{http://dx.doi.org/10.1103/PhysRevLett.89.011301}{{\em Phys.Rev.Lett.}
  {\bfseries 89} (2002) 011301},
  \href{http://arxiv.org/abs/nucl-ex/0204008}{{\ttfamily nucl-ex/0204008}}.

\bibitem{Fukuda:1998mi}
{\bfseries Super-Kamiokande} Collaboration, Y.~Fukuda {\em et~al.}, ``{Evidence
  for oscillation of atmospheric neutrinos},''
  \href{http://dx.doi.org/10.1103/PhysRevLett.81.1562}{{\em Phys.Rev.Lett.}
  {\bfseries 81} (1998) 1562--1567},
  \href{http://arxiv.org/abs/hep-ex/9807003}{{\ttfamily hep-ex/9807003}}.

\bibitem{McDonald:2016ixn}
A.~B. McDonald, ``{Nobel Lecture: The Sudbury Neutrino Observatory: Observation
  of flavor change for solar neutrinos},''
  \href{http://dx.doi.org/10.1103/RevModPhys.88.030502}{{\em Rev.Mod.Phys.}
  {\bfseries 88} no.~3, (2016) 030502}.

\bibitem{Kajita:2016cak}
T.~Kajita, ``{Nobel Lecture: Discovery of atmospheric neutrino oscillations},''
  \href{http://dx.doi.org/10.1103/RevModPhys.88.030501}{{\em Rev.Mod.Phys.}
  {\bfseries 88} no.~3, (2016) 030501}.

\bibitem{Eguchi:2002dm}
{\bfseries KamLAND} Collaboration, K.~Eguchi {\em et~al.}, ``{First results
  from KamLAND: Evidence for reactor anti-neutrino disappearance},''
  \href{http://dx.doi.org/10.1103/PhysRevLett.90.021802}{{\em Phys.Rev.Lett.}
  {\bfseries 90} (2003) 021802},
  \href{http://arxiv.org/abs/hep-ex/0212021}{{\ttfamily hep-ex/0212021}}.

\bibitem{Miranda:2000bi}
O.~Miranda, C.~Pena-Garay, T.~Rashba, V.~Semikoz, and J.~Valle, ``{The Simplest
  resonant spin flavor solution to the solar neutrino problem},''
  \href{http://dx.doi.org/10.1016/S0550-3213(00)00546-0}{{\em Nucl. Phys. B}
  {\bfseries 595} (2001) 360--380},
  \href{http://arxiv.org/abs/hep-ph/0005259}{{\ttfamily arXiv:hep-ph/0005259}}.

\bibitem{Miranda:2001hv}
O.~Miranda, C.~Pena-Garay, T.~Rashba, V.~Semikoz, and J.~Valle, ``{A
  Nonresonant dark side solution to the solar neutrino problem},''
  \href{http://dx.doi.org/10.1016/S0370-2693(01)01231-X}{{\em Phys. Lett. B}
  {\bfseries 521} (2001) 299--307},
  \href{http://arxiv.org/abs/hep-ph/0108145}{{\ttfamily arXiv:hep-ph/0108145}}.

\bibitem{Barranco:2002te}
J.~Barranco, O.~Miranda, T.~Rashba, V.~Semikoz, and J.~Valle, ``{Confronting
  spin flavor solutions of the solar neutrino problem with current and future
  solar neutrino data},''
  \href{http://dx.doi.org/10.1103/PhysRevD.66.093009}{{\em Phys.Rev.}
  {\bfseries D66} (2002) 093009},
  \href{http://arxiv.org/abs/hep-ph/0207326}{{\ttfamily hep-ph/0207326}}.

\bibitem{GonzalezGarcia:1998hj}
M.~Gonzalez-Garcia, M.~Guzzo, P.~Krastev, H.~Nunokawa, O.~Peres, V.~Pleitez,
  J.~Valle, and R.~Zukanovich~Funchal, ``{Atmospheric neutrino observations and
  flavor changing interactions},''
  \href{http://dx.doi.org/10.1103/PhysRevLett.82.3202}{{\em Phys. Rev. Lett.}
  {\bfseries 82} (1999) 3202--3205},
  \href{http://arxiv.org/abs/hep-ph/9809531}{{\ttfamily arXiv:hep-ph/9809531}}.

\bibitem{Guzzo:2001mi}
M.~Guzzo, P.~de~Holanda, M.~Maltoni, H.~Nunokawa, M.~Tortola, and J.~W.~F.
  Valle, ``{Status of a hybrid three neutrino interpretation of neutrino
  data},'' \href{http://dx.doi.org/10.1016/S0550-3213(02)00139-6}{{\em
  Nucl.Phys.} {\bfseries B629} (2002) 479--490},
  \href{http://arxiv.org/abs/hep-ph/0112310}{{\ttfamily hep-ph/0112310}}.

\bibitem{Miranda:2004nb}
O.~G. Miranda, M.~A. Tortola, and J.~W.~F. Valle, ``{Are solar neutrino
  oscillations robust?},''
  \href{http://dx.doi.org/10.1088/1126-6708/2006/10/008}{{\em JHEP} {\bfseries
  10} (2006) 008}, \href{http://arxiv.org/abs/hep-ph/0406280}{{\ttfamily
  hep-ph/0406280}}.

\bibitem{Esteban:2018ppq}
I.~Esteban, M.~Gonzalez-Garcia, M.~Maltoni, I.~Martinez-Soler, and J.~Salvado,
  ``{Updated Constraints on Non-Standard Interactions from Global Analysis of
  Oscillation Data},'' \href{http://dx.doi.org/10.1007/JHEP08(2018)180}{{\em
  JHEP} {\bfseries 08} (2018) 180},
  \href{http://arxiv.org/abs/1805.04530}{{\ttfamily arXiv:1805.04530
  [hep-ph]}}.

\bibitem{Dev:2019anc}
P.~B. Dev {\em et~al.}, ``{Neutrino Non-Standard Interactions: A Status
  Report},'' \href{http://dx.doi.org/10.21468/SciPostPhysProc.2.001}{{\em
  SciPost Phys.Proc.} {\bfseries 2} (2019) 001},
  \href{http://arxiv.org/abs/1907.00991}{{\ttfamily arXiv:1907.00991
  [hep-ph]}}.

\bibitem{Gariazzo:2015rra}
S.~Gariazzo, C.~Giunti, M.~Laveder, Y.~F. Li, and E.~M. Zavanin, ``{Light
  sterile neutrinos},''
  \href{http://dx.doi.org/10.1088/0954-3899/43/3/033001}{{\em J.Phys.G}
  {\bfseries 43} (2016) 033001},
  \href{http://arxiv.org/abs/1507.08204}{{\ttfamily arXiv:1507.08204
  [hep-ph]}}.

\bibitem{Gariazzo:2017fdh}
S.~Gariazzo, C.~Giunti, M.~Laveder, and Y.~F. Li, ``{Updated Global 3+1
  Analysis of Short-BaseLine Neutrino Oscillations},''
  \href{http://dx.doi.org/10.1007/JHEP06(2017)135}{{\em JHEP} {\bfseries 06}
  (2017) 135}, \href{http://arxiv.org/abs/1703.00860}{{\ttfamily
  arXiv:1703.00860 [hep-ph]}}.

\bibitem{Dentler:2018sju}
M.~Dentler, A.~Hern\'andez-Cabezudo, J.~Kopp, P.~A. Machado, M.~Maltoni,
  I.~Martinez-Soler, and T.~Schwetz, ``{Updated Global Analysis of Neutrino
  Oscillations in the Presence of eV-Scale Sterile Neutrinos},''
  \href{http://dx.doi.org/10.1007/JHEP08(2018)010}{{\em JHEP} {\bfseries 08}
  (2018) 010}, \href{http://arxiv.org/abs/1803.10661}{{\ttfamily
  arXiv:1803.10661 [hep-ph]}}.

\bibitem{Dentler:2017tkw}
M.~Dentler, A.~Hern\'{a}ndez-Cabezudo, J.~Kopp, M.~Maltoni, and T.~Schwetz,
  ``{Sterile Neutrinos or Flux Uncertainties? - Status of the Reactor
  Anti-Neutrino Anomaly},''
  \href{http://dx.doi.org/10.1007/JHEP11(2017)099}{{\em JHEP} {\bfseries 11}
  (2017) 099}, \href{http://arxiv.org/abs/1709.04294}{{\ttfamily
  arXiv:1709.04294 [hep-ph]}}.

\bibitem{Gariazzo:2018mwd}
S.~Gariazzo, C.~Giunti, M.~Laveder, and Y.~F. Li, ``{Model-Independent
  $\bar\nu_{e}$ Short-Baseline Oscillations from Reactor Spectral Ratios},''
  \href{http://dx.doi.org/10.1016/j.physletb.2018.04.057}{{\em Phys.Lett.}
  {\bfseries B782} (2018) 13--21},
  \href{http://arxiv.org/abs/1801.06467}{{\ttfamily arXiv:1801.06467
  [hep-ph]}}.

\bibitem{Diaz:2019fwt}
A.~Diaz, C.~Arg\"uelles, G.~Collin, J.~Conrad, and M.~Shaevitz, ``{Where Are We
  With Light Sterile Neutrinos?},''
  \href{http://dx.doi.org/10.1016/j.physrep.2020.08.005}{{\em Phys. Rept.}
  {\bfseries 884} (2020) 1--59},
  \href{http://arxiv.org/abs/1906.00045}{{\ttfamily arXiv:1906.00045
  [hep-ex]}}.

\bibitem{Giunti:2019aiy}
C.~Giunti and T.~Lasserre, ``{eV-scale Sterile Neutrinos},''
  \href{http://dx.doi.org/10.1146/annurev-nucl-101918-023755}{{\em
  Ann.Rev.Nucl.Part.Sci.} {\bfseries 69} no.~1, (2019) 163--190},
  \href{http://arxiv.org/abs/1901.08330}{{\ttfamily arXiv:1901.08330
  [hep-ph]}}.

\bibitem{Boser:2019rta}
S.~B\"oser, C.~Buck, C.~Giunti, J.~Lesgourgues, L.~Ludhova, S.~Mertens,
  A.~Schukraft, and M.~Wurm, ``{Status of Light Sterile Neutrino Searches},''
  \href{http://dx.doi.org/10.1016/j.ppnp.2019.103736}{{\em Prog. Part. Nucl.
  Phys.} {\bfseries 111} (2020) 103736},
  \href{http://arxiv.org/abs/1906.01739}{{\ttfamily arXiv:1906.01739
  [hep-ex]}}.

\bibitem{Cleveland:1998nv}
B.~Cleveland {\em et~al.}, ``{Measurement of the solar electron neutrino flux
  with the Homestake chlorine detector},''
  \href{http://dx.doi.org/10.1086/305343}{{\em Astrophys.J.} {\bfseries 496}
  (1998) 505--526}.

\bibitem{Kaether:2010ag}
F.~Kaether, W.~Hampel, G.~Heusser, J.~Kiko, and T.~Kirsten, ``{Reanalysis of
  the GALLEX solar neutrino flux and source experiments},''
  \href{http://dx.doi.org/10.1016/j.physletb.2010.01.030}{{\em Phys.Lett.B}
  {\bfseries 685} (2010) 47--54},
  \href{http://arxiv.org/abs/1001.2731}{{\ttfamily arXiv:1001.2731 [hep-ex]}}.

\bibitem{Abdurashitov:2009tn}
{\bfseries SAGE} Collaboration, J.~N. Abdurashitov {\em et~al.}, ``{Measurement
  of the solar neutrino capture rate with gallium metal. III: Results for the
  2002--2007 data-taking period},''
  \href{http://dx.doi.org/10.1103/PhysRevC.80.015807}{{\em Phys.Rev.C}
  {\bfseries 80} (2009) 015807},
  \href{http://arxiv.org/abs/0901.2200}{{\ttfamily arXiv:0901.2200 [nucl-ex]}}.

\bibitem{Bellini:2011rx}
G.~Bellini {\em et~al.}, ``{Precision measurement of the 7Be solar neutrino
  interaction rate in Borexino},''
  \href{http://dx.doi.org/10.1103/PhysRevLett.107.141302}{{\em Phys. Rev.
  Lett.} {\bfseries 107} (2011) 141302},
  \href{http://arxiv.org/abs/1104.1816}{{\ttfamily arXiv:1104.1816 [hep-ex]}}.

\bibitem{Bellini:2013lnn}
{\bfseries Borexino} Collaboration, G.~Bellini {\em et~al.}, ``{Final results
  of Borexino Phase-I on low energy solar neutrino spectroscopy},''
  \href{http://dx.doi.org/10.1103/PhysRevD.89.112007}{{\em Phys.Rev.D}
  {\bfseries 89} (2014) 112007},
  \href{http://arxiv.org/abs/1308.0443}{{\ttfamily arXiv:1308.0443 [hep-ex]}}.

\bibitem{Hosaka:2005um}
{\bfseries Super-Kamiokande} Collaboration, J.~Hosaka {\em et~al.}, ``{Solar
  neutrino measurements in super-Kamiokande-I},''
  \href{http://dx.doi.org/10.1103/PhysRevD.73.112001}{{\em Phys.Rev.D}
  {\bfseries 73} (2006) 112001},
  \href{http://arxiv.org/abs/hep-ex/0508053}{{\ttfamily arXiv:hep-ex/0508053
  [hep-ex]}}.

\bibitem{Cravens:2008aa}
{\bfseries Super-Kamiokande} Collaboration, J.~Cravens {\em et~al.}, ``{Solar
  neutrino measurements in Super-Kamiokande-II},''
  \href{http://dx.doi.org/10.1103/PhysRevD.78.032002}{{\em Phys.Rev.D}
  {\bfseries 78} (2008) 032002},
  \href{http://arxiv.org/abs/0803.4312}{{\ttfamily arXiv:0803.4312 [hep-ex]}}.

\bibitem{Abe:2010hy}
{\bfseries Super-Kamiokande} Collaboration, K.~Abe {\em et~al.}, ``{Solar
  neutrino results in Super-Kamiokande-III},''
  \href{http://dx.doi.org/10.1103/PhysRevD.83.052010}{{\em Phys.Rev.D}
  {\bfseries 83} (2011) 052010},
  \href{http://arxiv.org/abs/1010.0118}{{\ttfamily arXiv:1010.0118 [hep-ex]}}.

\bibitem{Nakano:PhD}
Y.~Nakano, ``{PhD Thesis, University of Tokyo}.''
  \url{http://www-sk.icrr.u-tokyo.ac.jp/sk/_pdf/articles/2016/doc_thesis_naknao.pdf},
  2016.

\bibitem{gioacchino_ranucci_2020_4134014}
G.~Ranucci, ``{First detection of solar neutrinos from CNO cycle with
  Borexino},'' June, 2020.
\newblock \url{https://doi.org/10.5281/zenodo.4134014}.

\bibitem{yasuhiro_nakajima_2020_4134680}
Y.~Nakajima, ``{Recent results and future prospects from Super- Kamiokande},''
  June, 2020.
\newblock \url{https://doi.org/10.5281/zenodo.4134680}.

\bibitem{Aharmim:2011vm}
{\bfseries SNO} Collaboration, B.~Aharmim {\em et~al.}, ``{Combined Analysis of
  all Three Phases of Solar Neutrino Data from the Sudbury Neutrino
  Observatory},'' \href{http://dx.doi.org/10.1103/PhysRevC.88.025501}{{\em
  Phys. Rev. C} {\bfseries 88} (2013) 025501},
  \href{http://arxiv.org/abs/1109.0763}{{\ttfamily arXiv:1109.0763 [nucl-ex]}}.

\bibitem{Vinyoles:2016djt}
N.~Vinyoles, A.~M. Serenelli, F.~L. Villante, S.~Basu, J.~Bergstrom, M.~C.
  Gonzalez-Garcia, M.~Maltoni, C.~Pe{\~n}a-Garay, and N.~Song, ``{A new
  Generation of Standard Solar Models},''
  \href{http://dx.doi.org/10.3847/1538-4357/835/2/202}{{\em Astrophys.J.}
  {\bfseries 835} (2017) 202},
  \href{http://arxiv.org/abs/1611.09867}{{\ttfamily arXiv:1611.09867
  [astro-ph.SR]}}.

\bibitem{Abe:2008aa}
{\bfseries KamLAND} Collaboration, S.~Abe {\em et~al.}, ``{Precision
  Measurement of Neutrino Oscillation Parameters with KamLAND},''
  \href{http://dx.doi.org/10.1103/PhysRevLett.100.221803}{{\em Phys. Rev.
  Lett.} {\bfseries 100} (2008) 221803},
  \href{http://arxiv.org/abs/0801.4589}{{\ttfamily arXiv:0801.4589 [hep-ex]}}.

\bibitem{Gando:2010aa}
{\bfseries KamLAND} Collaboration, A.~Gando {\em et~al.}, ``{Constraints on
  $\theta_{13}$ from A Three-Flavor Oscillation Analysis of Reactor
  Antineutrinos at KamLAND},''
  \href{http://dx.doi.org/10.1103/PhysRevD.83.052002}{{\em Phys.Rev.D}
  {\bfseries 83} (2011) 052002},
  \href{http://arxiv.org/abs/1009.4771}{{\ttfamily arXiv:1009.4771 [hep-ex]}}.

\bibitem{Gando:2013nba}
{\bfseries KamLAND} Collaboration, A.~Gando {\em et~al.}, ``{Reactor On-Off
  Antineutrino Measurement with KamLAND},''
  \href{http://dx.doi.org/10.1103/PhysRevD.88.033001}{{\em Phys. Rev. D}
  {\bfseries 88} no.~3, (2013) 033001},
  \href{http://arxiv.org/abs/1303.4667}{{\ttfamily arXiv:1303.4667 [hep-ex]}}.

\bibitem{Escrihuela:2009up}
F.~Escrihuela, O.~Miranda, M.~Tortola, and J.~W.~F. Valle, ``{Constraining
  nonstandard neutrino-quark interactions with solar, reactor and accelerator
  data},'' \href{http://dx.doi.org/10.1103/PhysRevD.80.129908}{{\em Phys. Rev.
  D} {\bfseries 80} (2009) 105009},
  \href{http://arxiv.org/abs/0907.2630}{{\ttfamily arXiv:0907.2630 [hep-ph]}}.
  {[Erratum: Phys.Rev.D 80, 129908 (2009)]}.

\bibitem{Coloma:2016gei}
P.~Coloma and T.~Schwetz, ``{Generalized mass ordering degeneracy in neutrino
  oscillation experiments},''
  \href{http://dx.doi.org/10.1103/PhysRevD.94.055005}{{\em Phys.Rev. D}
  {\bfseries 94} no.~5, (2016) 055005},
  \href{http://arxiv.org/abs/1604.05772}{{\ttfamily arXiv:1604.05772
  [hep-ph]}}. {[Erratum: Phys.Rev.D 95, 079903 (2017)]}.

\bibitem{Goswami:2004cn}
S.~Goswami and A.~Y. Smirnov, ``{Solar neutrinos and 1-3 leptonic mixing},''
  \href{http://dx.doi.org/10.1103/PhysRevD.72.053011}{{\em Phys. Rev. D}
  {\bfseries 72} (2005) 053011},
  \href{http://arxiv.org/abs/hep-ph/0411359}{{\ttfamily arXiv:hep-ph/0411359}}.

\bibitem{Bak:2018ydk}
{\bfseries RENO} Collaboration, G.~Bak {\em et~al.}, ``{Measurement of Reactor
  Antineutrino Oscillation Amplitude and Frequency at RENO},''
  \href{http://dx.doi.org/10.1103/PhysRevLett.121.201801}{{\em Phys.Rev.Lett.}
  {\bfseries 121} (2018) 201801},
  \href{http://arxiv.org/abs/1806.00248}{{\ttfamily arXiv:1806.00248
  [hep-ex]}}.

\bibitem{Adey:2018zwh}
{\bfseries Daya Bay} Collaboration, D.~Adey {\em et~al.}, ``{Measurement of
  electron antineutrino oscillation with 1958 days of operation at Daya Bay},''
  \href{http://dx.doi.org/10.1103/PhysRevLett.121.241805}{{\em Phys.Rev.Lett.}
  {\bfseries 121} (2018) 241805},
  \href{http://arxiv.org/abs/1809.02261}{{\ttfamily arXiv:1809.02261
  [hep-ex]}}.

\bibitem{Nunokawa:2005nx}
H.~Nunokawa, S.~J. Parke, and R.~Zukanovich~Funchal, ``{Another possible way to
  determine the neutrino mass hierarchy},''
  \href{http://dx.doi.org/10.1103/PhysRevD.72.013009}{{\em Phys. Rev.}
  {\bfseries D72} (2005) 013009},
  \href{http://arxiv.org/abs/hep-ph/0503283}{{\ttfamily arXiv:hep-ph/0503283
  [hep-ph]}}.

\bibitem{Hernandez-Cabezudo:2019qko}
J.~Hernandez-Cabezudo, S.~J. Parke, and S.-H. Seo, ``{Constraint on the Solar
  dm$^2$ from combined Daya Bay \& RENO data},''
  \href{http://dx.doi.org/10.1103/PhysRevD.100.113008}{{\em Phys.Rev. D}
  {\bfseries 100} (2019) 113008},
  \href{http://arxiv.org/abs/1905.09479}{{\ttfamily arXiv:1905.09479
  [hep-ex]}}.

\bibitem{Seo:2016uom}
{\bfseries RENO} Collaboration, S.~Seo {\em et~al.}, ``{Spectral Measurement of
  the Electron Antineutrino Oscillation Amplitude and Frequency using 500 Live
  Days of RENO Data},''
  \href{http://dx.doi.org/10.1103/PhysRevD.98.012002}{{\em Phys.Rev. D}
  {\bfseries 98} (2018) 012002},
  \href{http://arxiv.org/abs/1610.04326}{{\ttfamily arXiv:1610.04326
  [hep-ex]}}.

\bibitem{jonghee_yoo_2020_4123573}
J.~Yoo, ``Reno,'' June, 2020.
\newblock \url{https://doi.org/10.5281/zenodo.4123573}.

\bibitem{RENO:2015ksa}
{\bfseries RENO} Collaboration, J.~Choi {\em et~al.}, ``{Observation of Energy
  and Baseline Dependent Reactor Antineutrino Disappearance in the RENO
  Experiment},'' \href{http://dx.doi.org/10.1103/PhysRevLett.116.211801}{{\em
  Phys.Rev.Lett.} {\bfseries 116} (2016) 211801},
  \href{http://arxiv.org/abs/1511.05849}{{\ttfamily arXiv:1511.05849
  [hep-ex]}}.

\bibitem{An:2016srz}
{\bfseries Daya Bay} Collaboration, F.~P. An {\em et~al.}, ``{Improved
  Measurement of the Reactor Antineutrino Flux and Spectrum at Daya Bay},''
  \href{http://dx.doi.org/10.1088/1674-1137/41/1/013002}{{\em Chin.Phys.C}
  {\bfseries 41} (2017) 013002},
  \href{http://arxiv.org/abs/1607.05378}{{\ttfamily arXiv:1607.05378
  [hep-ex]}}.

\bibitem{An:2016ses}
{\bfseries Daya Bay} Collaboration, F.~P. An {\em et~al.}, ``{Measurement of
  electron antineutrino oscillation based on 1230 days of operation of the Daya
  Bay experiment},'' \href{http://dx.doi.org/10.1103/PhysRevD.95.072006}{{\em
  Phys.Rev.D} {\bfseries 95} (2017) 072006},
  \href{http://arxiv.org/abs/1610.04802}{{\ttfamily arXiv:1610.04802
  [hep-ex]}}.

\bibitem{Abe:2017aap}
{\bfseries Super-Kamiokande} Collaboration, K.~Abe {\em et~al.}, ``{Atmospheric
  neutrino oscillation analysis with external constraints in Super-Kamiokande
  I-IV},'' \href{http://dx.doi.org/10.1103/PhysRevD.97.072001}{{\em Phys.Rev.
  D} {\bfseries 97} (2018) 072001},
  \href{http://arxiv.org/abs/1710.09126}{{\ttfamily arXiv:1710.09126
  [hep-ex]}}.

\bibitem{Aartsen:2017nmd}
{\bfseries IceCube} Collaboration, M.~G. Aartsen {\em et~al.}, ``{Measurement
  of Atmospheric Neutrino Oscillations at 6-56 GeV with IceCube DeepCore},''
  \href{http://dx.doi.org/10.1103/PhysRevLett.120.071801}{{\em Phys.Rev.Lett.}
  {\bfseries 120} (2018) 071801},
  \href{http://arxiv.org/abs/1707.07081}{{\ttfamily arXiv:1707.07081
  [hep-ex]}}.

\bibitem{Aartsen:2019tjl}
M.~Aartsen {\em et~al.}, ``{Measurement of Atmospheric Tau Neutrino Appearance
  with IceCube DeepCore},''
  \href{http://dx.doi.org/10.1103/PhysRevD.99.032007}{{\em Phys.Rev. D}
  {\bfseries 99} (2019) 032007},
  \href{http://arxiv.org/abs/1901.05366}{{\ttfamily arXiv:1901.05366
  [hep-ex]}}.

\bibitem{Jiang:2019xwn}
{\bfseries Super-Kamiokande} Collaboration, M.~Jiang {\em et~al.},
  ``{Atmospheric Neutrino Oscillation Analysis With Improved Event
  Reconstruction in Super-Kamiokande IV},''
  \href{http://dx.doi.org/10.1093/ptep/ptz015}{{\em PTEP} {\bfseries 2019}
  (2019) 053F01}, \href{http://arxiv.org/abs/1901.03230}{{\ttfamily
  arXiv:1901.03230 [hep-ex]}}.

\bibitem{SKIV-tabs}
\url{http://www-sk.icrr.u-tokyo.ac.jp/sk/publications/data/sk.atm.data.release.tar.gz}.

\bibitem{Aartsen:2014yll}
{\bfseries IceCube} Collaboration, M.~G. Aartsen {\em et~al.}, ``{Determining
  neutrino oscillation parameters from atmospheric muon neutrino disappearance
  with three years of IceCube DeepCore data},''
  \href{http://dx.doi.org/10.1103/PhysRevD.91.072004}{{\em Phys.Rev.D}
  {\bfseries 91} (2015) 072004},
  \href{http://arxiv.org/abs/1410.7227}{{\ttfamily arXiv:1410.7227 [hep-ex]}}.

\bibitem{SampleA:IceCube-August-2019}
{\bfseries IceCube} Collaboration, ``{Three-year high-statistics neutrino
  oscillation samples}.''
  \url{https://icecube.wisc.edu/science/data/highstats_nuosc_3y}, 2019.

\bibitem{alex_himmel_2020_3959581}
{Alex Himmel}, ``{New Oscillation Results from the {NOvA} Experiment},'' Jul,
  2020.
\newblock \url{https://doi.org/10.5281/zenodo.3959581}.

\bibitem{patrick_dunne_2020_3959558}
{Patrick Dunne}, ``{Latest Neutrino Oscillation Results from {T2K}},'' Jul,
  2020.
\newblock \url{https://doi.org/10.5281/zenodo.3959558}.

\bibitem{Adamson:2014vgd}
{\bfseries MINOS} Collaboration, P.~Adamson {\em et~al.}, ``{Combined analysis
  of $\nu_{\mu}$ disappearance and $\nu_{\mu} \rightarrow \nu_{e}$ appearance
  in MINOS using accelerator and atmospheric neutrinos},''
  \href{http://dx.doi.org/10.1103/PhysRevLett.112.191801}{{\em Phys.Rev.Lett.}
  {\bfseries 112} (2014) 191801},
  \href{http://arxiv.org/abs/1403.0867}{{\ttfamily arXiv:1403.0867 [hep-ex]}}.

\bibitem{Ahn:2006zza}
{\bfseries K2K} Collaboration, M.~Ahn {\em et~al.}, ``{Measurement of Neutrino
  Oscillation by the K2K Experiment},''
  \href{http://dx.doi.org/10.1103/PhysRevD.74.072003}{{\em Phys.Rev.D}
  {\bfseries 74} (2006) 072003},
  \href{http://arxiv.org/abs/hep-ex/0606032}{{\ttfamily hep-ex/0606032}}.

\bibitem{Abe:2019ffx}
K.~Abe {\em et~al.}, ``{Search for Electron Antineutrino Appearance in a
  Long-baseline Muon Antineutrino Beam},''
  \href{http://dx.doi.org/10.1103/PhysRevLett.124.161802}{{\em Phys.Rev.Lett.}
  {\bfseries 124} (2020) 161802},
  \href{http://arxiv.org/abs/1911.07283}{{\ttfamily arXiv:1911.07283
  [hep-ex]}}.

\bibitem{Abe:2019vii}
K.~Abe {\em et~al.}, ``{Constraint on the Matter-Antimatter Symmetry-Violating
  Phase in Neutrino Oscillations},''
  \href{http://dx.doi.org/10.1038/s41586-020-2177-0}{{\em Nature} {\bfseries
  580} (2020) 339--344}, \href{http://arxiv.org/abs/1910.03887}{{\ttfamily
  arXiv:1910.03887 [hep-ex]}}.

\bibitem{Abe:2018wpn}
{\bfseries T2K} Collaboration, K.~Abe {\em et~al.}, ``{Search for CP Violation
  in Neutrino and Antineutrino Oscillations by the T2K Experiment with
  $2.2\times10^{21}$ Protons on Target},''
  \href{http://dx.doi.org/10.1103/PhysRevLett.121.171802}{{\em Phys. Rev.
  Lett.} {\bfseries 121} no.~17, (2018) 171802},
  \href{http://arxiv.org/abs/1807.07891}{{\ttfamily arXiv:1807.07891
  [hep-ex]}}.

\bibitem{NOvA:2018gge}
{\bfseries NOvA} Collaboration, M.~Acero {\em et~al.}, ``{New constraints on
  oscillation parameters from $\nu_e$ appearance and $\nu_\mu$ disappearance in
  the NOvA experiment},''
  \href{http://dx.doi.org/10.1103/PhysRevD.98.032012}{{\em Phys.Rev. D}
  {\bfseries 98} (2018) 032012},
  \href{http://arxiv.org/abs/1806.00096}{{\ttfamily arXiv:1806.00096
  [hep-ex]}}.

\bibitem{Acero:2019ksn}
{\bfseries NOvA} Collaboration, M.~Acero {\em et~al.}, ``{First measurement of
  neutrino oscillation parameters using neutrinos and antineutrinos by NOvA},''
  \href{http://dx.doi.org/10.1103/PhysRevLett.123.151803}{{\em Phys.Rev.Lett.}
  {\bfseries 123} (2019) 151803},
  \href{http://arxiv.org/abs/1906.04907}{{\ttfamily arXiv:1906.04907
  [hep-ex]}}.

\bibitem{Huber:2004ka}
P.~Huber, M.~Lindner, and W.~Winter, ``{Simulation of long-baseline neutrino
  oscillation experiments with GLoBES (General Long Baseline Experiment
  Simulator)},'' \href{http://dx.doi.org/10.1016/j.cpc.2005.01.003}{{\em
  Comput. Phys. Commun.} {\bfseries 167} (2005) 195},
  \href{http://arxiv.org/abs/hep-ph/0407333}{{\ttfamily arXiv:hep-ph/0407333
  [hep-ph]}}.

\bibitem{Huber:2007ji}
P.~Huber, J.~Kopp, M.~Lindner, M.~Rolinec, and W.~Winter, ``{New features in
  the simulation of neutrino oscillation experiments with GLoBES 3.0: General
  Long Baseline Experiment Simulator},''
  \href{http://dx.doi.org/10.1016/j.cpc.2007.05.004}{{\em Comput. Phys.
  Commun.} {\bfseries 177} (2007) 432--438},
  \href{http://arxiv.org/abs/hep-ph/0701187}{{\ttfamily arXiv:hep-ph/0701187
  [hep-ph]}}.

\bibitem{Adamson:2013whj}
{\bfseries MINOS} Collaboration, P.~Adamson {\em et~al.}, ``{Measurement of
  Neutrino and Antineutrino Oscillations Using Beam and Atmospheric Data in
  MINOS},'' \href{http://dx.doi.org/10.1103/PhysRevLett.110.251801}{{\em Phys.
  Rev. Lett.} {\bfseries 110} no.~25, (2013) 251801},
  \href{http://arxiv.org/abs/1304.6335}{{\ttfamily arXiv:1304.6335 [hep-ex]}}.

\bibitem{Adamson:2013ue}
{\bfseries MINOS} Collaboration, P.~Adamson {\em et~al.}, ``{Electron neutrino
  and antineutrino appearance in the full MINOS data sample},''
  \href{http://dx.doi.org/10.1103/PhysRevLett.110.171801}{{\em Phys. Rev.
  Lett.} {\bfseries 110} no.~17, (2013) 171801},
  \href{http://arxiv.org/abs/1301.4581}{{\ttfamily arXiv:1301.4581 [hep-ex]}}.

\bibitem{Aliu:2004sq}
{\bfseries K2K} Collaboration, E.~Aliu {\em et~al.}, ``{Evidence for muon
  neutrino oscillation in an accelerator-based experiment},''
  \href{http://dx.doi.org/10.1103/PhysRevLett.94.081802}{{\em Phys. Rev. Lett.}
  {\bfseries 94} (2005) 081802},
  \href{http://arxiv.org/abs/hep-ex/0411038}{{\ttfamily arXiv:hep-ex/0411038}}.

\bibitem{An:2015jdp}
{\bfseries JUNO} Collaboration, F.~An {\em et~al.}, ``{Neutrino Physics with
  JUNO},'' \href{http://dx.doi.org/10.1088/0954-3899/43/3/030401}{{\em J.Phys.
  G} {\bfseries 43} no.~3, (2016) 030401},
  \href{http://arxiv.org/abs/1507.05613}{{\ttfamily arXiv:1507.05613
  [physics.ins-det]}}.

\bibitem{Tortola:2020ncu}
M.~A. T\'ortola, G.~Barenboim, and C.~A. Ternes, ``{CPT and CP, an entangled
  couple},'' \href{http://dx.doi.org/10.1007/JHEP07(2020)155}{{\em JHEP}
  {\bfseries 07} (2020) 155}, \href{http://arxiv.org/abs/2005.05975}{{\ttfamily
  arXiv:2005.05975 [hep-ph]}}.

\bibitem{Kelly:2020fkv}
K.~J. Kelly, P.~A. Machado, S.~J. Parke, Y.~F. Perez~Gonzalez, and
  R.~Zukanovich-Funchal, ``{Back to (Mass-)Square(d) One: The Neutrino Mass
  Ordering in Light of Recent Data},''
  \href{http://arxiv.org/abs/2007.08526}{{\ttfamily arXiv:2007.08526
  [hep-ph]}}.

\bibitem{Audren:2012wb}
B.~Audren, J.~Lesgourgues, K.~Benabed, and S.~Prunet, ``{Conservative
  Constraints on Early Cosmology: an illustration of the Monte Python
  cosmological parameter inference code},''
  \href{http://dx.doi.org/10.1088/1475-7516/2013/02/001}{{\em JCAP} {\bfseries
  02} (2013) 001}, \href{http://arxiv.org/abs/1210.7183}{{\ttfamily
  arXiv:1210.7183 [astro-ph.CO]}}.

\bibitem{Brinckmann:2018cvx}
T.~Brinckmann and J.~Lesgourgues, ``{MontePython 3: boosted MCMC sampler and
  other features},'' \href{http://dx.doi.org/10.1016/j.dark.2018.100260}{{\em
  Phys.Dark Univ.} {\bfseries 24} (2019) 100260},
  \href{http://arxiv.org/abs/1804.07261}{{\ttfamily arXiv:1804.07261
  [astro-ph.CO]}}.

\bibitem{Heavens:2017afc}
A.~Heavens {\em et~al.}, ``{Marginal Likelihoods from Monte Carlo Markov
  Chains},'' \href{http://arxiv.org/abs/1704.03472}{{\ttfamily arXiv:1704.03472
  [stat.CO]}}.

\bibitem{Handley:2015fda}
W.~J. Handley, M.~P. Hobson, and A.~N. Lasenby, ``{PolyChord: nested sampling
  for cosmology},'' \href{http://dx.doi.org/10.1093/mnrasl/slv047}{{\em
  Mon.Not.Roy.Astron.Soc.} {\bfseries 450} no.~1, (2015) L61--L65},
  \href{http://arxiv.org/abs/1502.01856}{{\ttfamily arXiv:1502.01856
  [astro-ph.CO]}}.

\bibitem{Handley:2015aa}
W.~J. Handley, M.~P. Hobson, and A.~N. Lasenby, ``{PolyChord: next-generation
  nested sampling},'' \href{http://dx.doi.org/10.1093/mnras/stv1911}{{\em
  Monthly Notices of the Royal Astronomical Society} {\bfseries 453} no.~4,
  (Jun, 2015) 4384}, \href{http://arxiv.org/abs/1506.00171}{{\ttfamily
  1506.00171}}.

\bibitem{Gariazzo:2018pei}
S.~Gariazzo, M.~Archidiacono, P.~F. de~Salas, O.~Mena, C.~A. Ternes, and
  M.~T{\'o}rtola, ``{Neutrino masses and their ordering: Global Data, Priors
  and Models},'' \href{http://dx.doi.org/10.1088/1475-7516/2018/03/011}{{\em
  JCAP} {\bfseries 03} (2018) 011},
  \href{http://arxiv.org/abs/1801.04946}{{\ttfamily arXiv:1801.04946
  [hep-ph]}}.

\bibitem{deSalas:2018bym}
P.~F. De~Salas, S.~Gariazzo, O.~Mena, C.~A. Ternes, and M.~T{\'o}rtola,
  ``{Neutrino Mass Ordering from Oscillations and Beyond: 2018 Status and
  Future Prospects},'' \href{http://dx.doi.org/10.3389/fspas.2018.00036}{{\em
  Front.Astron.Space Sci.} {\bfseries 5} (2018) 36},
  \href{http://arxiv.org/abs/1806.11051}{{\ttfamily arXiv:1806.11051
  [hep-ph]}}.

\bibitem{Giunti:2007ry}
C.~Giunti and C.~W. Kim, {\em {Fundamentals of Neutrino Physics and
  Astrophysics}}.
\newblock 2007.

\bibitem{Aker:2019uuj}
{\bfseries KATRIN} Collaboration, M.~Aker {\em et~al.}, ``{An improved upper
  limit on the neutrino mass from a direct kinematic method by KATRIN},''
  \href{http://dx.doi.org/10.1103/PhysRevLett.123.221802}{{\em Phys.Rev.Lett.}
  {\bfseries 123} (2019) 221802},
  \href{http://arxiv.org/abs/1909.06048}{{\ttfamily arXiv:1909.06048
  [hep-ex]}}.

\bibitem{Kraus:2004zw}
C.~Kraus {\em et~al.}, ``{Final results from phase II of the Mainz neutrino
  mass search in tritium beta decay},''
  \href{http://dx.doi.org/10.1140/epjc/s2005-02139-7}{{\em Eur.Phys.J.C}
  {\bfseries 40} (2005) 447--468},
  \href{http://arxiv.org/abs/hep-ex/0412056}{{\ttfamily arXiv:hep-ex/0412056
  [hep-ex]}}.

\bibitem{Aseev:2012zz}
V.~Aseev {\em et~al.}, ``{Measurement of the electron antineutrino mass in
  tritium beta decay in the Troitsk nu-mass experiment},''
  \href{http://dx.doi.org/10.1134/S1063778812030027}{{\em Phys.Atom.Nucl.}
  {\bfseries 75} (2012) 464--478}.

\bibitem{Huang:2019tdh}
G.-y. Huang, W.~Rodejohann, and S.~Zhou, ``{Effective neutrino masses in KATRIN
  and future tritium beta-decay experiments},''
  \href{http://dx.doi.org/10.1103/PhysRevD.101.016003}{{\em Phys.Rev. D}
  {\bfseries 101} (2020) 016003},
  \href{http://arxiv.org/abs/1910.08332}{{\ttfamily arXiv:1910.08332
  [hep-ph]}}.

\bibitem{Schechter:1981bd}
J.~Schechter and J.~W.~F. Valle, ``{Neutrinoless Double beta Decay in SU(2) x
  U(1) Theories},'' \href{http://dx.doi.org/10.1103/PhysRevD.25.2951}{{\em
  Phys.Rev.D} {\bfseries 25} (1982) 2951}.

\bibitem{Schechter:1980gr}
J.~Schechter and J.~W.~F. Valle, ``{Neutrino Masses in SU(2) x U(1)
  Theories},'' \href{http://dx.doi.org/10.1103/PhysRevD.22.2227}{{\em
  Phys.Rev.D} {\bfseries 22} (1980) 2227}.

\bibitem{Schechter:1980gk}
J.~Schechter and J.~Valle, ``{Neutrino Oscillation Thought Experiment},''
  \href{http://dx.doi.org/10.1103/PhysRevD.23.1666}{{\em Phys.Rev.} {\bfseries
  D23} (1981) 1666}.

\bibitem{Rodejohann:2011vc}
W.~Rodejohann and J.~Valle, ``{Symmetrical Parametrizations of the Lepton
  Mixing Matrix},'' \href{http://dx.doi.org/10.1103/PhysRevD.84.073011}{{\em
  Phys.Rev.} {\bfseries D84} (2011) 073011},
  \href{http://arxiv.org/abs/1108.3484}{{\ttfamily arXiv:1108.3484 [hep-ph]}}.

\bibitem{DellOro:2016tmg}
S.~Dell'Oro, S.~Marcocci, M.~Viel, and F.~Vissani, ``{Neutrinoless double beta
  decay: 2015 review},'' \href{http://dx.doi.org/10.1155/2016/2162659}{{\em
  Adv.High Energy Phys.} {\bfseries 2016} (2016) 2162659},
  \href{http://arxiv.org/abs/1601.07512}{{\ttfamily arXiv:1601.07512
  [hep-ph]}}.

\bibitem{Agostini:2019hzm}
{\bfseries GERDA} Collaboration, M.~Agostini {\em et~al.}, ``{Probing Majorana
  neutrinos with double-$\beta$ decay},''
  \href{http://dx.doi.org/10.1126/science.aav8613}{{\em Science} {\bfseries
  365} (2019) 1445}, \href{http://arxiv.org/abs/1909.02726}{{\ttfamily
  arXiv:1909.02726 [hep-ex]}}.

\bibitem{Adams:2019jhp}
{\bfseries CUORE} Collaboration, D.~Adams {\em et~al.}, ``{Improved Limit on
  Neutrinoless Double-Beta Decay in $^{130}$Te with CUORE},''
  \href{http://dx.doi.org/10.1103/PhysRevLett.124.122501}{{\em Phys.Rev.Lett.}
  {\bfseries 124} (2020) 122501},
  \href{http://arxiv.org/abs/1912.10966}{{\ttfamily arXiv:1912.10966
  [nucl-ex]}}.

\bibitem{KamLAND-Zen:2016pfg}
{\bfseries KamLAND-Zen} Collaboration, A.~Gando {\em et~al.}, ``{Search for
  Majorana Neutrinos near the Inverted Mass Hierarchy Region with
  KamLAND-Zen},'' \href{http://dx.doi.org/10.1103/PhysRevLett.117.109903}{{\em
  Phys.Rev.Lett.} {\bfseries 117} no.~8, (2016) 082503},
  \href{http://arxiv.org/abs/1605.02889}{{\ttfamily arXiv:1605.02889
  [hep-ex]}}. {[Addendum: Phys. Rev. Lett.117,no.10,109903(2016)]}.

\bibitem{Caldwell:2017mqu}
A.~Caldwell, A.~Merle, O.~Schulz, and M.~Totzauer, ``{A Global Bayesian
  Analysis of Neutrino Mass Data},''
  \href{http://dx.doi.org/10.1103/PhysRevD.96.073001}{{\em Phys.Rev. D}
  {\bfseries 96} (2017) 073001},
  \href{http://arxiv.org/abs/1705.01945}{{\ttfamily arXiv:1705.01945
  [hep-ph]}}.

\bibitem{Vergados:2016hso}
J.~Vergados, H.~Ejiri, and F.~{\v S}imkovic, ``{Neutrinoless double beta decay
  and neutrino mass},'' \href{http://dx.doi.org/10.1142/S0218301316300071}{{\em
  Int.J.Mod.Phys.} {\bfseries E25} (2016) 1630007},
  \href{http://arxiv.org/abs/1612.02924}{{\ttfamily arXiv:1612.02924
  [hep-ph]}}.

\bibitem{Gariazzo:2018meg}
S.~Gariazzo and O.~Mena, ``{Cosmology-marginalized approaches in Bayesian model
  comparison: the neutrino mass as a case study},''
  \href{http://dx.doi.org/10.1103/PhysRevD.99.021301}{{\em Phys.Rev. D}
  {\bfseries 99} (2019) 021301},
  \href{http://arxiv.org/abs/1812.05449}{{\ttfamily arXiv:1812.05449
  [astro-ph.CO]}}.

\bibitem{Lattanzi:2017ubx}
M.~Lattanzi and M.~Gerbino, ``{Status of neutrino properties and future
  prospects - Cosmological and astrophysical constraints},''
  \href{http://dx.doi.org/10.3389/fphy.2017.00070}{{\em Front.in Phys.}
  {\bfseries 5} (2018) 70}, \href{http://arxiv.org/abs/1712.07109}{{\ttfamily
  arXiv:1712.07109 [astro-ph.CO]}}.

\bibitem{Akrami:2018vks}
{\bfseries Planck} Collaboration, N.~Aghanim {\em et~al.}, ``{Planck 2018
  results. I. Overview and the cosmological legacy of Planck},''
  \href{http://dx.doi.org/10.1051/0004-6361/201833880}{{\em Astron. Astrophys.}
  {\bfseries 641} (2020) A1}, \href{http://arxiv.org/abs/1807.06205}{{\ttfamily
  arXiv:1807.06205 [astro-ph.CO]}}.

\bibitem{Aghanim:2018eyx}
{\bfseries Planck} Collaboration, N.~Aghanim {\em et~al.}, ``{Planck 2018
  results. VI. Cosmological parameters},''
  \href{http://dx.doi.org/10.1051/0004-6361/201833910}{{\em Astron. Astrophys.}
  {\bfseries 641} (2020) A6}, \href{http://arxiv.org/abs/1807.06209}{{\ttfamily
  arXiv:1807.06209 [astro-ph.CO]}}.

\bibitem{Aghanim:2019ame}
{\bfseries Planck} Collaboration, N.~Aghanim {\em et~al.}, ``{Planck 2018
  results. V. CMB power spectra and likelihoods},''
  \href{http://dx.doi.org/10.1051/0004-6361/201936386}{{\em Astron. Astrophys.}
  {\bfseries 641} (2020) A5}, \href{http://arxiv.org/abs/1907.12875}{{\ttfamily
  arXiv:1907.12875 [astro-ph.CO]}}.

\bibitem{Aghanim:2018oex}
{\bfseries Planck} Collaboration, N.~Aghanim {\em et~al.}, ``{Planck 2018
  results. VIII. Gravitational lensing},''
  \href{http://dx.doi.org/10.1051/0004-6361/201833886}{{\em Astron. Astrophys.}
  {\bfseries 641} (2020) A8}, \href{http://arxiv.org/abs/1807.06210}{{\ttfamily
  arXiv:1807.06210 [astro-ph.CO]}}.

\bibitem{Beutler:2011hx}
F.~Beutler, C.~Blake, M.~Colless, D.~H. Jones, L.~Staveley-Smith, L.~Campbell,
  Q.~Parker, W.~Saunders, and F.~Watson, ``{The 6dF Galaxy Survey: Baryon
  Acoustic Oscillations and the Local Hubble Constant},''
  \href{http://dx.doi.org/10.1111/j.1365-2966.2011.19250.x}{{\em
  Mon.Not.Roy.Astron.Soc.} {\bfseries 416} (2011) 3017--3032},
  \href{http://arxiv.org/abs/1106.3366}{{\ttfamily arXiv:1106.3366
  [astro-ph.CO]}}.

\bibitem{Ross:2014qpa}
A.~J. Ross, L.~Samushia, C.~Howlett, W.~J. Percival, A.~Burden, and M.~Manera,
  ``{The clustering of the SDSS DR7 main Galaxy sample - I. A 4~per cent
  distance measure at $z~=~0.15$},''
  \href{http://dx.doi.org/10.1093/mnras/stv154}{{\em Mon.Not.Roy.Astron.Soc.}
  {\bfseries 449} no.~1, (2015) 835--847},
  \href{http://arxiv.org/abs/1409.3242}{{\ttfamily arXiv:1409.3242
  [astro-ph.CO]}}.

\bibitem{Alam:2016hwk}
{\bfseries BOSS} Collaboration, S.~Alam {\em et~al.}, ``{The clustering of
  galaxies in the completed SDSS-III Baryon Oscillation Spectroscopic Survey:
  cosmological analysis of the DR12 galaxy sample},''
  \href{http://dx.doi.org/10.1093/mnras/stx721}{{\em Mon.Not.Roy.Astron.Soc.}
  {\bfseries 470} (2017) 2617--2652},
  \href{http://arxiv.org/abs/1607.03155}{{\ttfamily arXiv:1607.03155
  [astro-ph.CO]}}.

\bibitem{Moresco:2016mzx}
M.~Moresco, L.~Pozzetti, A.~Cimatti, R.~Jimenez, C.~Maraston, L.~Verde,
  D.~Thomas, A.~Citro, R.~Tojeiro, and D.~Wilkinson, ``{A 6\% measurement of
  the Hubble parameter at $z\sim0.45$: direct evidence of the epoch of cosmic
  re-acceleration},''
  \href{http://dx.doi.org/10.1088/1475-7516/2016/05/014}{{\em JCAP} {\bfseries
  05} no.~05, (2016) 014}, \href{http://arxiv.org/abs/1601.01701}{{\ttfamily
  arXiv:1601.01701 [astro-ph.CO]}}.

\bibitem{Scolnic:2017caz}
D.~M. Scolnic {\em et~al.}, ``{The Complete Light-curve Sample of
  Spectroscopically Confirmed SNe Ia from Pan-STARRS1 and Cosmological
  Constraints from the Combined Pantheon Sample},''
  \href{http://dx.doi.org/10.3847/1538-4357/aab9bb}{{\em Astrophys. J.}
  {\bfseries 859} no.~2, (2018) 101},
  \href{http://arxiv.org/abs/1710.00845}{{\ttfamily arXiv:1710.00845
  [astro-ph.CO]}}.

\bibitem{Riess:2019cxk}
A.~G. Riess, S.~Casertano, W.~Yuan, L.~M. Macri, and D.~Scolnic, ``{Large
  Magellanic Cloud Cepheid Standards Provide a 1\% Foundation for the
  Determination of the Hubble Constant and Stronger Evidence for Physics beyond
  $\Lambda$CDM},'' \href{http://dx.doi.org/10.3847/1538-4357/ab1422}{{\em
  Astrophys.J.} {\bfseries 876} (2019) 85},
  \href{http://arxiv.org/abs/1903.07603}{{\ttfamily arXiv:1903.07603
  [astro-ph.CO]}}.

\bibitem{Lesgourgues:2013bra}
J.~Lesgourgues and T.~Tram, ``{Fast and accurate CMB computations in non-flat
  FLRW universes},''
  \href{http://dx.doi.org/10.1088/1475-7516/2014/09/032}{{\em JCAP} {\bfseries
  09} no.~09, (2014) 032}, \href{http://arxiv.org/abs/1312.2697}{{\ttfamily
  arXiv:1312.2697 [astro-ph.CO]}}.

\bibitem{Blas:2011rf}
D.~Blas, J.~Lesgourgues, and T.~Tram, ``{The Cosmic Linear Anisotropy Solving
  System (CLASS) II: Approximation schemes},''
  \href{http://dx.doi.org/10.1088/1475-7516/2011/07/034}{{\em JCAP} {\bfseries
  07} (2011) 034}, \href{http://arxiv.org/abs/1104.2933}{{\ttfamily
  arXiv:1104.2933 [astro-ph.CO]}}.

\bibitem{Lesgourgues:2011re}
J.~Lesgourgues, ``{The Cosmic Linear Anisotropy Solving System (CLASS) I:
  Overview},'' \href{http://arxiv.org/abs/1104.2932}{{\ttfamily arXiv:1104.2932
  [astro-ph.IM]}}.

\bibitem{Astone:1999wp}
P.~Astone and G.~D'Agostini, ``{Inferring the intensity of Poisson processes at
  the limit of the detector sensitivity (with a case study on gravitational
  wave burst search)},'' {\em Annals Phys.} ,
  \href{http://arxiv.org/abs/hep-ex/9909047}{{\ttfamily hep-ex/9909047}}.

\bibitem{DAgostini:2000edp}
G.~D'Agostini, ``{Confidence limits: What is the problem? Is there the
  solution?},'' in {\em {Workshop on confidence limits, CERN, Geneva,
  Switzerland, 17-18 Jan 2000: Proceedings}}, pp.~3--27.
\newblock 2000.
\newblock \href{http://arxiv.org/abs/hep-ex/0002055}{{\ttfamily
  arXiv:hep-ex/0002055 [hep-ex]}}.

\bibitem{DAgostini:2003}
G.~D'Agostini, \href{http://dx.doi.org/10.1142/5262}{{\em {Bayesian Reasoning
  in Data Analysis}}}.
\newblock {WORLD} {SCIENTIFIC}, Jun, 2003.
\newblock \url{{https://doi.org/10.1142%2F5262}}.

\bibitem{Gariazzo:2019xhx}
S.~Gariazzo, ``{Constraining power of open likelihoods, made
  prior-independent},''
  \href{http://dx.doi.org/10.1140/epjc/s10052-020-8126-0}{{\em Eur. Phys. J. C}
  {\bfseries 80} no.~6, (2020) 552},
  \href{http://arxiv.org/abs/1910.06646}{{\ttfamily arXiv:1910.06646
  [astro-ph.CO]}}.

\bibitem{Vagnozzi:2018jhn}
S.~Vagnozzi, S.~Dhawan, M.~Gerbino, K.~Freese, A.~Goobar, and O.~Mena,
  ``{Constraints on the sum of the neutrino masses in dynamical dark energy
  models with $w(z) \geq -1$ are tighter than those obtained in
  $\Lambda$CDM},'' \href{http://dx.doi.org/10.1103/PhysRevD.98.083501}{{\em
  Phys.Rev.D} {\bfseries 98} (2018) 083501},
  \href{http://arxiv.org/abs/1801.08553}{{\ttfamily arXiv:1801.08553
  [astro-ph.CO]}}.

\bibitem{Vagnozzi:2017ovm}
S.~Vagnozzi, E.~Giusarma, O.~Mena, K.~Freese, M.~Gerbino, S.~Ho, and
  M.~Lattanzi, ``{Unveiling $\nu$ secrets with cosmological data: neutrino
  masses and mass hierarchy},''
  \href{http://dx.doi.org/10.1103/PhysRevD.96.123503}{{\em Phys.Rev. D}
  {\bfseries 96} (2017) 123503},
  \href{http://arxiv.org/abs/1701.08172}{{\ttfamily arXiv:1701.08172
  [astro-ph.CO]}}.

\bibitem{Giusarma:2016phn}
E.~Giusarma, M.~Gerbino, O.~Mena, S.~Vagnozzi, S.~Ho, and K.~Freese,
  ``{Improvement of cosmological neutrino mass bounds},''
  \href{http://dx.doi.org/10.1103/PhysRevD.94.083522}{{\em Phys.Rev.D}
  {\bfseries 94} no.~8, (2016) 083522},
  \href{http://arxiv.org/abs/1605.04320}{{\ttfamily arXiv:1605.04320
  [astro-ph.CO]}}.

\end{thebibliography}

\end{document}